\documentclass[iop]{emulateapj}
\usepackage{amsmath}
\usepackage{physics}
\usepackage{lineno}
\usepackage{multirow}
\usepackage{hyperref}
\usepackage{xcolor}
\usepackage{pgfplotstable,tabularx,booktabs}
\usepackage{lipsum, babel}
\pgfplotsset{compat=1.18}

\definecolor{darkgreen}{RGB}{0,100,0}

\shorttitle{Fast Reconnection in the Atmospheres of the Sun and Protoplanetary Disks}
\shortauthors{Fulvia Pucci, K.A.P. Singh, Uma Gorti, Marco Velli, Neal J.\ Turner, M. E. Innocenti}
\begin{document}
\title{Applications of Fast Magnetic Reconnection Models to the Atmospheres of the Sun and Protoplanetary Disks}

\author{Fulvia Pucci\altaffilmark{1}, K. Alkendra P.\ Singh\altaffilmark{2,3}, Uma Gorti\altaffilmark{4,5}, Marco Velli\altaffilmark{6}, Neal J.\ Turner\altaffilmark{1}, Disha Varshney\altaffilmark{2}, Maria Elena\ Innocenti\altaffilmark{7}} 
\email{fulvia.e.pucci@jpl.nasa.gov}
\affil{\altaffilmark{1}Jet Propulsion Laboratory, California Institute of Technology, 4800 Oak Grove Drive, Pasadena, California 91109, USA}
\affil{\altaffilmark{2}Plasma Astrophysics Research Laboratory, Department of Physics, Institute of Science, BHU, Varanasi 221005, India}
\affil{\altaffilmark{3}Astronomical Observatory, Graduate School of Science, Kyoto University , Yamashina, Kyoto 607-8471, Japan}
\affil{\altaffilmark{4}Department of Earth, Planetary, and Space Sciences, UCLA, 595 Charles Young Drive East, Los Angeles, California 90095, USA}
\affil{\altaffilmark{5}NASA Ames Research Center, MS 245-1, PO Box 1, Moffett Field, California 94035, USA}
\affil{\altaffilmark{6}SETI Institute, 339 Bernardo Ave Suite 200, Mountain View, California 94043, USA}
\affil{\altaffilmark{7}Institut f\"ur Theoretische Physik, Ruhr-Universit\"at Bochum, Bochum, Germany}

\begin{abstract}
Partially-ionized plasmas consist of charged and neutral particles whose mutual collisions modify magnetic reconnection compared with the fully-ionized case. The collisions alter the rate and locations of the magnetic dissipation heating and the distribution of energies among the particles accelerated into the non-thermal tail. We examine the collisional regimes for the onset of fast reconnection in two environments: the partially-ionized layers of the solar atmosphere and the protoplanetary disks that are the birthplaces for planets around young stars. In both these environments, magnetic nulls readily develop into resistive current sheets in the regime where the charged and neutral particles are fully coupled by collisions, but the current sheets quickly break down under the ideal tearing instability. The current sheets collapse repeatedly, forming magnetic islands at successively smaller scales, till they enter a collisionally-decoupled regime where the magnetic energy is rapidly turned into heat and charged-particle kinetic energy. Small-scale, decoupled fast reconnection in the solar atmosphere may lead to preferential heating and energization of ions and electrons that escape into the corona.  In protoplanetary disks such reconnection causes localized heating in the atmospheric layers that produce much of the infrared atomic and molecular line emission observed with the Spitzer and James Webb Space Telescopes.
\end{abstract}

\section{Introduction}

Magnetic reconnection plays a key role in the solar atmosphere, from the photosphere, chromosphere to filaments, and prominences, as well as in the interstellar medium with its star-forming molecular clouds, and the planet-forming protoplanetary discs that orbit young stars (e.g.\ \citealt{Ballester:2018, 2020RSPSA.47690867N, 2022NatRP...4..263J} and references therein) and many other environments.  Reconnection in stars and accretion disks contributes to the development of their hot coronae, the dynamo regeneration of their magnetic fields, and the launching of their supersonic winds (e.g.\ \citealt{ZweibelYamada:2009, Yamadaetal:2010, Yamada_2022} and references therein).  In many of these environments, the plasma is only partially ionized.  The substantial neutral fraction alters the rates of both reconnection \citep{Zweibel:1989,PhysRevLett.122.015101,Puccietal:2020} and the instabilities that can disrupt the current sheets (CSs) where reconnection takes place \citep{SolerBallester2022}.  Thus the partially-ionized nature of the plasma must be part of attempts to understand these systems.

Through the solar photosphere and chromosphere the plasma density varies sharply with height and the ionization degree of hydrogen varies from about $\chi\sim 10^{-4}$ in the photosphere to $\chi\sim1$ at the top of the chromosphere (\citealt{1981ApJS...45..635V,1993ApJ...406..319F}).  A wide variety of dynamical events which characterize these regions are attributed, at least in part, to reconnection:
chromospheric jets (\citealt{2007Sci...318.1591S,2009ApJ...707L..37L,2012A&A...543A...6M,2014Sci...346A.315T}),
Ellerman Bombs (\citealt{2006ApJ...643.1325F,2015ApJ...798...19N}), and
type II white light flares (e.g., \citealt{1999ApJ...512..454D}).
Recently, improvements in the angular resolution of solar telescopes have allowed many small-scale events in the low solar atmosphere to be observed \citep{2015ApJ...798L..11Y,2016NatCo...711837X}.
High-temperature compact bright points, possible signatures of magnetic reconnection, have UV counterparts frequently observed with the Interface Region Imaging Spectrograph (IRIS; \citealt{2014Sci...346C.315P}). 

Protoplanetary disks (PPDs) consist of gas and dust orbiting young stars.  They form as a consequence of angular momentum conservation during the collapse under self-gravity of dense cores within interstellar molecular gas clouds. They therefore are a universal step in forming stars \citep{1987ARA&A..25...23S} and provide the raw materials from which planets form \citep{2003PASP..115..965E}.

The ionization in PPDs' surface layers comes mainly from photons emitted by the central star in the X-ray \citep{Igea_1999} and UV bands \citep{2011ApJ...735....8P}.  The disk midplane is optically-thick to these photons, but remains ionized at least weakly, thanks to cosmic rays \citep{2014A&A...571A..33P} and the decay of radioactive isotopes \citep{2009ApJ...690...69U}.  The ionization fraction $\chi$ ranges from $<10^{-14}$ near the midplane \citep{10.3389/fspas.2018.00039} to $>10^{-4}$ in the disk atmosphere, where far-UV photons ionize most of the carbon atoms \citep{2011ApJ...735....8P}.  Midplane temperatures are high enough for thermal ionization in the fraction of an au nearest the star \citep{2015ApJ...811..156D}.

Magnetic fields threading the disks play fundamental roles in the evolution and dispersal of the planet-forming material.  The fields readily slip through the gas in the weakly-ionized parts of the disk interior and are better-coupled to the gas in the more-strongly-ionized surface layers.
Reconnection can affect many aspects of the planet-forming environment, including the saturation amplitude of the turbulence driven by magneto-rotational instability or MRI \citep{1999ApJ...515..776S,2001ApJ...561L.179S}, the formation of gas rings that may determine planets' initial locations \citep{2018MNRAS.477.1239S, 2019MNRAS.484..107S} and the localized heating needed to form the chondrules common in primitive meteorites \citep{2013ApJ...767L...2M}.  Magnetic forces also launch winds from the disk atmosphere near the plasma $\beta=1$ surface where the dynamics transitions from gas-dominated to magnetically-dominated \citep{2015ApJ...801...84G, 2020ApJ...896..126G, 2023ASPC..534..567P, 2022A&A...667A..17M}.  Flows near this surface, dragging the magnetic field footpoints at the wind source can generate CSs leading to magnetic reconnection.  While the processes listed above may be understood through modeling they remain difficult to observe directly, though there is near-term potential for chemical and thermodynamic signatures of reconnection to be detected through molecular near- and mid-infrared spectroscopy with the James Webb Space Telescope (JWST).

The idea that magnetic reconnection operates similarly across the solar atmosphere, protoplanetary disks, and other astrophysical contexts, comes from the fact that any CS forming in these environments is subject to the tearing instability \citep{FKR:1963,Loureiro:2007}, leading to magnetic reconnection.  Reconnection follows a longer period over which flows in the plasma store energy locally, in macroscopic, coherent magnetic structures.  Once reconnection begins, the stored magnetic energy is quickly converted into plasma heating and particle acceleration.  The conditions for triggering such fast reconnection can be understood using the ideal tearing model (IT, \citealt{PucciVelli:2014}).  Under this model, the CS first thins gradually as reconnection proceeds slowly. Then, once the CS aspect ratio (length over thickness) reaches a critical threshold, further evolution is dominated by a fast tearing mode.  This model explains numerical MHD  results from \citet{2015ApJ...806..131L,Teneranietal:2015b,Landi17, Huangetal:2017} and plasma kinetic results from \citet{DelSartoetal:2016, Puccietal:2017,2020ApJ...902..142S}.  The variety of scales involved and the universality of the process suggest that, in the partially ionized case \citep{ Zweibel:1989,Puccietal:2020}, there are three regimes of reconnection: (1) coupled ion and neutral dynamics, (2) weakly coupled and (3) decoupled dynamics. These regimes are separated by the thickness of the CS being larger or smaller than characteristic scales, defined by the parameters of the system and, in particular, by collision frequencies of ions with neutrals. In \citet{Puccietal:2020} the onset of fast magnetic reconnection generated by the tearing instability is discussed for each regime.

In this work we apply the model to reconnection in the lower solar atmosphere and protoplanetary disks.
In Sec.~\ref{FastRec:Summary} we summarize the processes involved in the onset of fast magnetic reconnection.
In Sec.~\ref{PIPSummary} we generalize to partially-ionized plasmas.
In Sec.~\ref{cascade-fractal} we present analytical calculations of the transitions between the reconnection regimes due to recursive processes.
In Sec.~\ref{ReynoldApp} we introduce the spatial and temporal scales involved in the reconnection process in the solar chromosphere and PPDs.
In Sec.~\ref{Sect:Sun} we apply the model to the solar atmosphere, examining the length scales at which magnetic reconnection can be observed in the lower, partially-ionized layers and comparing them with the observable range of scales.
In Sec.~\ref{Sect:Disks} we carry out similar calculations for PPD atmospheres, contrast the reconnection regimes with the magnetic diffusion regimes commonly considered in this context,
and finally estimate the local heating resulting from the dissipation of the magnetic fields.
A summary and conclusions are in Sec.~\ref{Sect:Conclusions}.

\section{Onset of fast magnetic reconnection, a summary.}
\label{FastRec:Summary}
In this section, we summarize the onset of the tearing mode instability, leading to fast magnetic reconnection in a resistive, fully-ionized plasma. For simplicity we describe the process starting from a 2D Harris CS \citep{1962NCim...23..115H}{, though the analysis is also valid when there is an out of plane component of the magnetic field, i.e.\ a guide field, that may also vary so as to make the equilibrium force-free -- see e.g.\ Eq.~(5) and Eq.~(6) in \citealt{2015ApJ...806..131L}}.  The scale of the variation of the equilibrium magnetic field determines the initial CS thickness $a$.

The linear stability (for incompressible fluctuations) does not depend on the presence or absence of a magnetic field in the direction orthogonal to the reconnection plane ($z$ or $\hat{k}$) and whether the equilibrium is force-free or pressure balanced \citep{2020ApJ...902..142S}. 
At large Lundquist numbers (low collision regimes) two regions define the solution structure of the perturbed magnetohydrodynamics (MHD) equations: a boundary layer of thickness 2$\delta$ around the center ($y = 0$) of the CS, where $\delta$ is defined so it separates resistive inner regions $y<\delta$ from outer regions where magnetic diffusivity $\eta_O$ due to resistivity and growth rate of the tearing instability may be neglected. 
From here on, barred quantities are normalized to the CS \emph{thickness} $a$. $\bar\tau_A = a/v_A$ is the Alfv\'en crossing time using the Alfv\'en speed $v_A$, $\bar k = ka=2\pi n/\mathcal{L}$ where $n$ is the wavenumber so that the minimum wavevector available in the system is $\bar k_{min} = 2\pi/\mathcal{L}$, for some system lengthscale $\mathcal{L}$. The Lundquist number $\bar S = \bar\tau_R/\bar\tau_A = a\,v_A/\eta_O$, $\bar\tau_R = a^2/\eta_O$ is the Ohmic diffusion time over the thickness $a$. 
The dependence of the maximum growth rate of the tearing instability $\gamma_{\rm max}$ on the Lundquist number (Ohmic magnetic diffusivity) can be expressed as 
\begin{equation}
\label{maxtear}
\gamma_{\rm max} \bar \tau_A \sim \bar S^{ -\frac{1}{2}}\,,\,\,\,\frac{\delta_{\rm max}}{a} \sim \bar S^{ -\frac{1}{4}}\,,\,\,\,
k_{\rm max} a\sim \bar S^{-\frac{1}{4}}.
\end{equation}
where the subscript ``max'' indicates the maximum growth rate available, corresponding to the wavevector $k_{\rm max}$ (see e.g.\ \citealt{DelSartoetal:2016} for the Harris CS). The expression for $\gamma_{\rm max}$ in Eq.~\eqref{maxtear} suggests that when the Lundquist number is large, as in most astrophysical plasmas,
the growth rate becomes negligible and the process inefficient. 
We also notice that the system has an intrinsic length scale $L$, limiting the length of the CS (e.g.\ the length of a magnetic loop in the solar corona, the pressure scale height in a disk atmosphere etc..). 
This suggested the relation for the ``ideal'' tearing instability \cite{PucciVelli:2014}, i.e.\ for an instability where the growth rate survives independently of the Lundquist number in the ideal limit $S\rightarrow \infty$. Rescaling the dispersion relation to the CS length rather than the thickness, the maximum growth rate of the tearing instability (corresponding to the fastest growing mode) reads:
\begin{equation}
\label{harrisdisp}
\gamma \tau_A \sim S^{-\frac{1}{2}}\left({\frac{a}{L}}\right)^{-\frac{3}{2}}.
\end{equation}
where we dropped the subscript ``max'' to simplify the notation.
For an inverse aspect ratio varying as $a/L\sim S^{-\alpha}$, any $\alpha<1/3$ leads to growth rates diverging in the ideal limit, while any $\alpha>1/3$ leads to growth rates tending to zero as the Lundquist number grows \citep{PucciVelli:2014,Teneranietal:2015b}.
This result is general: additional effects such as viscosity \citep{Teneranietal:2015} and Hall current \citep{Puccietal:2017} result in different scalings for the critical aspect ratio at which fast reconnection is triggered. It also extends to situations where non-collisional terms in Ohm's law break the frozen-in conditions \citep[e.g.][]{DelSartoetal:2016,mallet_2020}.

\section{Reconnection in partially ionized plasmas.}
\label{PIPSummary}
In partially-ionized plasmas, the electron-neutral, ion-neutral and electron-ion collisions cause an Ohmic-type diffusion of the magnetic field.  With three different species undergoing collisions, such as electrons, ions, and neutrals, a single-fluid description leads to an appropriately modified magnetic induction equation \citep[][eq.~28]{2007Ap&SS.311...35W}:
\begin{eqnarray}
&&\dfrac{\partial B}{\partial t}= 
\nabla \times ({\bf v}\times {\bf B})- \nabla \times [ \eta_O(\nabla \times {\bf B})\nonumber\\[2ex]
&&+\eta_H(\nabla \times {\bf B})\times {\hat{b}}+\eta_{AD} (\nabla \times {\bf B})_{\perp}]
\label{adimentionalind}
\end{eqnarray}
where $\hat{b}={\bf B}/|{\bf B}|$, and $\eta_H$, and $\eta_{AD}$ are respectively the Hall, and ambipolar diffusivities (AD).  
The velocity is here the velocity of the neutral fluid, and the coefficients are calculated neglecting electron pressure and other forces associated with the ionized components that are assumed to be negligible.
If the only charge carriers are ions and electrons the Ohmic diffusivity becomes:
\begin{equation}
\label{etaohm}
\eta_{O}=\dfrac{c^2}{\omega^2_{pe}}(\nu_{ei}+\nu_{en}) +\dfrac{c^2}{\omega^2_{pi}}\nu_{in}
\end{equation}
where $\omega_{pe}$ is the electron plasma frequency, the electron-ion, ion-neutral and electron-neutral collision frequencies are respectively $\nu_{ei,in,en}$ and $c$ is the speed of light. In the lower solar atmosphere helium and hydrogen are by far the most abundant species \citep{1993ApJ...406..319F} and collisions between neutrals and ions are negligible so the last term in Eq.~\eqref{etaohm} can be dropped.
For protoplanetary disks, usually the electron ion-collisions are negligible and, under this assumption, in \cite{2007Ap&SS.311...35W}, a prescription for the calculation of non-ideal coefficients is provided as a function of the local magnetic field, ions and neutrals collision frequencies, masses and abundances. \cite{2007Ap&SS.311...35W} describes three regimes in which each of the terms in Eq.~\eqref{etaohm} dominates the diffusion process in protoplanetary disks, where the main charge carriers may vary from dust grains near the midplane, to electrons and $H^+$ ions in the upper atmospheric layers (see Sec.~\ref{Sect:Disks}). 
The ambipolar diffusivity can be written as (see e.g.\ \citealt{2020RSPSA.47690867N}):
\begin{equation}
\label{etaad}
\eta_{AD}=\dfrac{{|\bf  B}|^2}{|\mu_0|}
\left( \dfrac{\rho_n}{\rho}\right)^2(\rho_i\nu_{in}+\rho_e\nu_{en})^{-1}
\end{equation}
where $\rho=\rho_i+\rho_n$ is the total mass density, $\rho_i$, $\rho_n$ and $\rho_e$ are respectively the ion, neutral and electron mass density; $\mu_0$ is the vacuum permeability. Assuming quasi neutrality ($n_e\sim n_i$), this expression is the same as in \cite{2007Ap&SS.311...35W}, where $\rho\sim\rho_n$.
Usually diffusion processes are slow compared to the advection time and the local heating they provide depends on the actual value of the coefficients. In the IT scenario magnetic reconnection supported by Ohmic diffusivity can occur in a time comparable to the magnetic field advection $\tau_{rec} \sim \tau_A$. 
The Hall effect can modify the reconnection scenario
when the scale of the equilibrium magnetic field gradient $a$, generating the CS, is $a \sim d_i=c/\omega_{pi}$, where $d_i$ is the ion inertial length and $\omega_{pi}$ the ion plasma frequency \citep{Terasawa:1983}. In particular for modifications of the IT model, see \cite{Puccietal:2017} for the linear theory, where the critical aspect ratio depends on the ion inertial length while reconnection still proceeds at the fastest speed in the system, and \cite{Shietal19}
for nonlinear simulations.
In the linear analysis of reconnecting instabilities the effect of non-ideal terms is assessed by their relative weight compared to the first advective term on the right hand side of Eq.~\eqref{adimentionalind}. 
For the AD to become relevant specifically for reconnection dynamics,
\begin{equation}
\label{ADimportant}
a\le v^2_{Ai}/(v_A \nu_{in}).
\end{equation}

{Note that AD does not violate the flux conservation associated with electron and ion motion. It appears in Ohm's law only because the advective term is written for the frame of the neutral fluid. The assumption is that collisions partly couple neutrals to ions so the neutrals participate indirectly in the reconnection process.
In the discussion that follows, we assume instability sets in on an initial state where ions and neutrals move with the same velocity.
We include AD and its effect on the linear instability.
However AD may also play a role in the formation of favorable initial conditions with CSs \citep{BrandenburgZweibel:1994,1999ApJ...511..193V,1965RvPP....1..205B}.  AD-driven CS steepening may be inhibited by guide fields \citep{2020RSPSA.47690867N}, though this does not affect the stability directly, in the sense that for a given CS thickness the presence or absence of a guide field does not change the sheet's stability.}

The relative importance of the different non-ideal effects in the partially ionized environments of the solar atmosphere and protoplanetary accretion disks are discussed in detail below.

\subsection{Reconnection: the tearing mode equations.}
\label{FastPIP}

Recently \cite{Puccietal:2020} extended the instability calculations presented by
\citet{Zweibel:1989} to the full resistive tearing mode dispersion relation, and from there obtained the trigger conditions for IT in the partially ionized case. In the latter work, there is no drift of the ions with respect to neutrals in the equilibrium. The current carriers in the starting equilibrium CS, considered to be force-free, are therefore the electrons only. In the linearized equation for incompressible perturbations
the ions and neutrals are collisionally coupled but the equations are then resolved in terms of the ion flow. Electron-neutral collisions are not taken into account in Ohm's law, again written in terms of the perturbed ion flow, though we consider the electron-neutral collisions in our calculation of the diffusive coefficients here. Compared to Eq.~\eqref{adimentionalind}, the \citet{Zweibel:1989} model neglects the Hall effect, while AD does not appear explicitly because of the different choice of variables in Ohm's law (ion flow vs neutral flow), though the corresponding couplings, namely ion neutral momentum transfers, are included via the modified ion momentum equation.  For the different forms of Ohm's law we refer to
\cite{MalyshkinZweibel:2011,MurphyLukin:2015}.

The Hall effect was considered by \cite{Terasawa:1983,Puccietal:2017,2020ApJ...902..142S}, from which we derive where the Hall effect plays a role in our calculations and discuss its effects on the triggering of fast magnetic reconnnection. 

In \citet{Puccietal:2020, Zweibel:1989} the linearized momentum and induction equations are written for the ion velocity and  vector potentials (respectively $\phi\, ,\psi$) as (primes denote derivatives with respect 
to the $a$-scaled variable $y/a$):
\begin{eqnarray}
\label{piptear}
(\gamma \bar \tau_{Ai})^2\left( 1+\dfrac{\nu_{in}}{\gamma+\nu_{ni}}\right)(\phi''- \bar k^2\phi)=\nonumber\\
-F(\psi''-\bar k^2\psi)+F''\psi
\\
\psi=\bar kF\phi+ \frac{1}{\bar S\gamma \bar \tau_{Ai}} (\psi''-\bar k^2\psi),\nonumber
\end{eqnarray}
where $\bar S$ is defined using the Alfv\'en speed calculated with the ion mass density.
We want to stress again that barred quantities will be normalized to the CS thickness, so that, for example $\bar \tau_{Ai}=a/v_{Ai}$ is the Alfv\'en time calculated with the ion density, $\gamma$ is the tearing growth rate associated with a mode with wave vector $\bar k=ka$ along the equilibrium magnetic field. 
We calculate the collision frequencies assuming binary elastic (energy and momentum conservation) collisions between electrons and neutrals so that $\nu_{ni} = \dfrac{n_im_i}{n_nm_n} \ \nu_{in} \Rightarrow \nu_{ni} < \nu_{in}$ at most heights in the solar atmosphere (see Tab.~1 in \citealt{Singhetal:2015}) and in protoplanetary disks. Note that the opposite limit $\nu_{ni} \gg \nu_{in}$ leads to the standard tearing of a completely ionized plasma.
Following \citet{Zweibel:1989} we may redefine a starred Alfv\'en time and Lundquist number 
\begin{eqnarray}
\label{Zweibelpres}
\bar \tau_{Ai} (1+\frac{\nu_{in}}{\gamma +
\nu_{ni}})^{1/2}&:=& \bar \tau_{Ai} f_M^{1/2}   \rightarrow \bar{\tau}_A^{*},\\
\label{SandSStar}
\bar S^*&:=& \bar S \frac{\bar \tau_{Ai}}{\bar{\tau}_A^{*}}= \bar{S}  f_M^{-1/2}.
\end{eqnarray}
\noindent Inserting $\bar{\tau}_A^{*}$ into Eqs.~\eqref{piptear} and substituting $\bar S \bar \tau_{Ai}$ with $\bar S^*\bar{\tau}_A^{*}$ and $\gamma\bar \tau_{Ai}$ with $\gamma \bar{\tau}^*_A$,
the tearing mode equations regain their standard form, so that all the properties of the dispersion relation discussed previously now apply to the starred quantities.
We are going to analyze the maximum growth rate of the tearing instability Eq.~\eqref{maxtear}, because the fastest growing mode is the most relevant in the context of triggering fast magnetic reconnection in natural plasmas causing an efficient energy conversion.
In particular, from Eq.~\eqref{maxtear} we have that $\gamma \bar{\tau}_A^{*}$ follows the same scaling with $\bar S^*$ as in the standard tearing theory:
\begin{eqnarray}
\label{standardgrowth}
\gamma \bar{\tau}^*_A \sim (\bar S^*)^{-1/2} \Rightarrow \gamma \bar \tau_{Ai}\sim (\bar S)^{-1/2}\left(\dfrac{\bar \tau_{Ai}}{\bar{\tau}^*_A}\right)^{1/2}.
\end{eqnarray}
\noindent  When the growth rate is negligible compared to both collision frequencies, the factor $f_M^{1/2}$ becomes
\begin{equation}
\label{idio1}
f_M^{1/2}=(1+\frac{\nu_{in}}{\nu_{ni}})^{1/2}=(1+\frac{\rho_{n}}
{\rho_{i}})^{1/2}=\left(\frac{\rho}{\rho_{i}}\right)^{1/2}.
\end{equation}

\subsection{Reconnection regimes.}
\begin{figure*}
 \includegraphics[width=180mm]{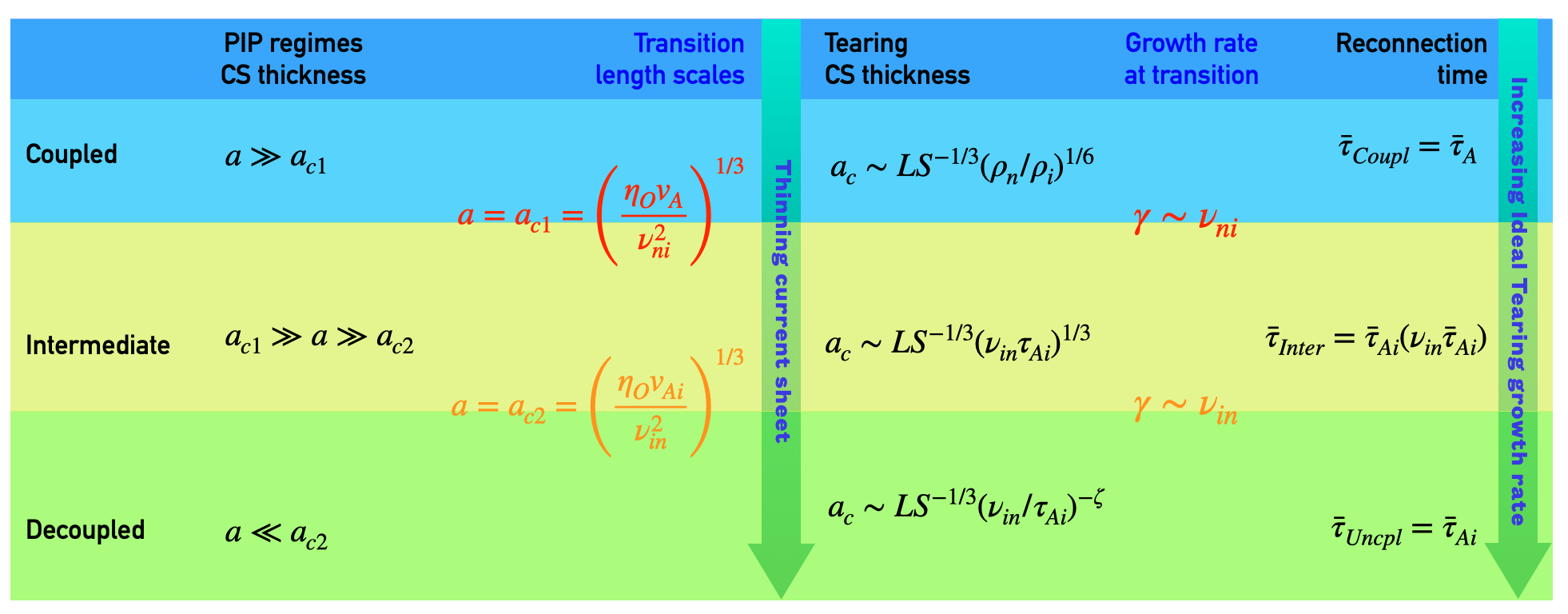}
\caption{Diagram summarizing magnetic reconnection regimes.  From top to bottom, the CS thickness decreases and reconnection happens faster.  CSs can shrink also through the recursive operation of tearing instability. 
At left on each row is the CS thickness $a$ for fast reconnection, relative to the critical length scales $a_{c1}$ and $a_{c2}$.  At right are the critical IT CS thickness for fast reconnection and the timescale for the IT instability.  In the coupled regime, the characteristic reconnection time is the Alfve\'n time including the density of all species.  In the decoupled regime, the characteristic time is the Alfve\'n time using the density of the ions alone.  The critical thickness then depends weakly on $\nu_{in} / \tau_{Ai}$ to the exponent $\zeta$.  This results in reconnection speeding up as we move from the top of the diagram to the bottom.  A single CS can pass through all three regimes while thinning, until either the magnetic field is consumed or the recursive tearing ends because the trigger conditions are no longer met.}
 \label{RegTransition}
\end{figure*}

With the two ion-neutral collision frequencies, two intrinsic length scales are introduced into the resistive MHD equations,
$a_{c1,c2}$ defined as \citep{Puccietal:2020}
\begin{equation}
\label{CriticalScalesPartial}
a_{c1,c2} = \left({\eta_m v_{A,Ai}}/{\nu_{ni,in}^2}\right)^{1/3}.
\end{equation}
If the thickness of the CS $a\gg a_{c1}$ fast reconnection onset kicks in in a regime where the neutrals and ion dynamics is coupled. For $a_{c1}\ge a \ge a_{c2}$ fast reconnection onset occurs in the intermediate regime. Finally when $a<a_{c2}$ fast reconnection dynamics involves ions only, with neutrals not really noticing. In Fig.~\ref{RegTransition} (left two columns) we summarized the regimes (divided by colors) and the transition scales (labelled in red and orange), showing that from the coupled to the decoupled regime we are transitioning to thinner CS. Ultimately, the plasma parameters and dynamics determine the regime in which fast reconnection onset occur.

In Fig.~\ref{RegTransition} right three columns, we summarize the results for the onset of fast magnetic reconnection from \citet{Puccietal:2020}.  In particular, we report the critical CS thickness $a_c$ for each regime, the values of the growth rate at the transitions between regimes (red and orange colors) and the timescale over which the instability proceeds in each regime once the critical thickness is reached.
The latter depends on the ion-neutral collision frequencies, and in particular in the fully decoupled regime has a weak dependence on $\nu_{in} / \tau_{Ai}$ expressed via the exponent $\zeta$
\noindent (non barred quantities are normalized with $L$ so that $S=L v_{Ai}/\eta_O$).
The growth rate of the tearing instability and so the speed at which reconnection proceeds in the fully coupled regime will be of the order of $\tau_A$, where the Alfv\'en speed is calculated with the total (ions+neutrals density).
In particular when $a_c\ll a_{c1,c2}$ the time-scale becomes the shortest ion-only Alfv\'en time, 
$\tau_{Ai}=\bar \tau_{Ai} a/L$. 
{To compare the trigger conditions found in \cite{Puccietal:2020} with those arising when including explicit ambipolar, Hall and Pedersen terms in Ohm's law \cite{2007Ap&SS.311...35W} and differential equilibrium flows between neutrals and ions, a linear study of the tearing instability should be carried out within a three fluid framework, where the dynamics of the positive and negative charge carriers and of the neutrals are solved for the IT mode.  However, the main AD effects on magnetic reconnection, i.e.\ the consequences of ion - neutral drift on reconnecting mode growth rates are captured by the critical scales of \cite{Puccietal:2020}}.

\section{Energy cascade towards smaller scales}
\label{cascade-fractal}

On the basis of the above discussion, one may outline the behavior of the tearing instability in a simple CS that is slowly thinning.  We shall assume the tearing mode develops on the global, ion-neutral coupled, Alfv\'enic time-scale, since this often proves to be the case for the environments discussed in this work.  Modeling work has shown that a recursive reconnection regime may appear \citep{ShibataTanuma:2001,Teneranietal:2015b,2015PhPl...22j0706S,2017ApJ...849...75H}, where CSs form at successively smaller scales. This suggests a transition to the intermediate and then fully decoupled regime as the sheets' thicknesses decrease, accelerating the nonlinear evolution of the tearing mode, and so of the energy conversion.

The natural thinning of the CS results from nonlinear processes driven by the self-attraction of parallel currents and the repulsion of oppositely directed currents. Considering a simple 2D magnetic configuration as in \cite{Teneranietal:2015b}, the reconnecting magnetic field generates an inflow-outflow pattern at the x-point. These flows tend to collapse the x-point into a new sheet that is stretched by the outflow but being kept at the initial thickness by the inflow. Once the sheet lengthens sufficiently, it itself is unstable to reconnection on a faster timescale, breaking the initially macroscopic CS into ever smaller reconnecting regions. Each then becomes further unstable accelerating the process until a turbulent regime, with heating and acceleration, is obtained (Fig.~2 in \citealt{Teneranietal:2015b}, Fig.2 in \citealt{2015PhPl...22j0706S}, Fig.~\citealt{2017ApJ...849...75H}).
 As discussed by \citet{ShibataTanuma:2001,Singhetal:2019}, different recursive reconnection models predict different numbers of plasmoids at the $n$-th stage of the reconnection cascade and so different number of CSs separating plasmoids (regions of close magnetic field lines, see the figures and references mentioned above). The process continues down to smaller scales until energy is dissipated, by relevant microscopic processes; for the fully ionized corona, for example, these are collective kinetic effects. In partially-ionized plasmas, if a reconnection event is initiated at scales at which ions and neutrals are coupled, see the blue area in the diagram in Fig.~\ref{RegTransition}, as sheets keep shrinking, they will become unstable to secondary IT instability in the intermediate or fully decoupled regimes.
Information on plasmoid scaling and energy transfer to the small scales can be translated into spectral features \citep{2022JPlPh..88e1501S}, and not only can help distinguish one model prediction to another, but might also result as an observable feature of the reconnnection layer.

\subsection{Transition from the coupled to the Intermediate regime.}
Understanding the energy cascade at smaller scales requires finding a relation between the $n_{th}$ recursive reconnection stage properties and the $n_{th}-1$ parent reconnecting process. As discussed in \citet{Teneranietal:2016}, we do not make any assumption as in e.g.\ \citet{ShibataTanuma:2001, JiandDaughton:2011}. We follow instead \citet{Teneranietal:2015b} where it is empirically observed that the characteristics of the $n_{th}$ CS thickness is related to the $n_{th}-1$ CS length by $a_n/L_{n-1}\sim S^{-1/2}$, because the upstream magnetic field is seen to be swept into the reforming x-points and remains of the same intensity as the background at each level (see e.g.\ \citealt{2022JPlPh..88e1501S}). For sake of simplicity let's consider the second step of the recursive collapse, then
\begin{equation}
a_1/L_0\sim S^{-1/2},
\label{recursivecoll2}
\end{equation} 
where we labelled with subscript $0$ the initial macroscopic unstable CS length, with subscript $1$ the parameters of the secondary reconnecting sheets, and assumed we are in the fully coupled ion-neutral regime. Later in this work we will show this hypothesis is satisfied for a relatively large range of CS lengths in the lower solar atmosphere as well as in protoplanetary disks. In the partially ionized case $\delta$ corresponding to the maximum growth rate is given by $\delta/ a \sim \bar{S}^{*-1/4}=(\tau_D/\bar{\tau}_A^{*})^{-1/4}=S^{-1/4} f_M^{-1/8}$, where again $S$ is the Lunquist number calculated with Alfv\'en speed based on the ion mass density only. As discussed in \citet{Puccietal:2020}, since $f_M$ is invariant for the IT rescaling, the solution is the same as for the classic IT with corrections depending on the regime, $\delta / L =S^{-1/2} f_M^{-1/8}$. Based on the assumptions above, Eq.~\eqref{recursivecoll2} becomes 
\begin{equation}
a_1/L_0\sim S^{-1/2}(\rho_n/\rho_i)^{-1/8}
\end{equation}
This means that if the CS thickness at the $n=1$ recursive reconnection stage $a_1<a_{c1}$, the triggering condition for the tearing instability at this stage will be the one described in Fig.~\ref{RegTransition} for the intermediate regime, i.e.\ $a_c\sim LS^{-1/3}(\nu_{in}/\tau_{Ai})^{1/3}$. This occurs when 
\begin{equation}
a_{1} < a_{c1}.\nonumber
\end{equation}
The inequality above gives
\begin{eqnarray}
\tau_{A}/\tau_{ni} < (S\ \rho/\rho_i)^{1/4}
\label{secondaryTrigger2}
\end{eqnarray}
where $\tau_{ni}\sim 1/\nu_{ni}$ as the collision time neutral-ion.

As we will show in the next sections, in a partially ionized plasma with a significant neutral population, Eq.~\eqref{secondaryTrigger2} is often satisfied and secondary reconnection will occur in the intermediate or decoupled regime, even if the dynamics ultimately depend on the thickness and so length of the CS and the local Lundquist number.

\section{{Advection scales in Solar and disk atmospheres.}}
\label{ReynoldApp}

{In this section we address the characteristic lengths and temporal scale variations of the environments in question, in order to calculate magnetic Reynolds numbers necessary to determine the critical thicknesses of fast magnetic reconnection.}
{The reason we focus on the Reynolds number is the following. Once a CS has formed and become unstable to fast magnetic reconnection thanks to resistivity and neutral collisions breaking flux-freezing, the sheet's critical thickness is the key parameter to understand the role of other non-ideal effects in the reconnection process. The critical thickness, estimated as shown in Fig.~\ref{RegTransition} can be then compared to the lengths calculated in Sec.~\ref{FastRec:Summary} to find whether the Hall effect and AD alter reconnection's onset. AD is included in our model and affects the reconnection regimes in which a degree of coupling between ions and neutrals is present. The Hall effect has scales defined by the ion inertial length $d_i$ and affects reconnection at kinetic scales, where collisions between neutrals and ions are generally negligible. In this paper we will discuss in detail when corrections due to the Hall effect and AD are relevant, even if they are not expected to significantly change the conclusions.  The picture could be further refined using three-fluid calculations to obtain corrections to the CS thickness explicitly proportional to $\eta_{AD}$.  However, such corrections would be small since the neutrals' role in the ion and electron dynamics is already included here.}

We adopt the following definition for the magnetic Reynolds number:
\begin{equation}
\label{standardLundquist}
R_m=\frac{L v}{ \eta_O}.
\end{equation} 
where $L$ is the characteristic macroscopic lengthscale of the system and $v$ is the convective speed found in the first term on the right-hand side of Eq.~\eqref{adimentionalind}.  The Lundquist number $S$ is obtained by replacing $v$ with $v_A$.
Eq.~\eqref{harrisdisp} for fully ionized plasmas and Eq.~\eqref{idio1} for partially ionized plasmas tell us the efficiency of reconnection due to the tearing instability  depends on the magnetic Reynolds number.
For the model in \cite{Puccietal:2020} this dependence reflects on the critical scale of which fast magnetic reconnection- and so efficient magnetic energy conversion- can be achieved. In order to discuss the reconnection dynamics it is then important to understand how to estimate the magnetic Reynolds numbers. In this section we will discuss the (temporal and spatial) scales involved in convecting and diffusing the magnetic field in the solar chromosphere and photosphere, as well as in protoplanetary disks. A similar discussion holds for any other plasma environment. 

\subsection{Magnetic Reynolds number in the solar atmosphere.}
\label{ReynoldsSolar}
{The solar photosphere and chromosphere are gravitationally stratified.  This means the characteristic length scale $L$ of reconnection in these layers is limited to at most the smaller of the pressure scale height $H_p$ and the typical transverse convective scale, defined by granulation at $\sim1000$~km.  Since $H_p$ is the smaller of the two, $L < H_p$.
In the solar corona in contrast, the gravitational scale height is much greater than the transverse and magnetic field scales, the plasma $\beta$ ($\simeq c_{\rm s}^2/v_A^2$ ) is very small, and gravity can often be neglected.
In the photosphere and chromosphere then, the pressure scale height defines the maximum vertical coherent scale achievable so that the appropriate magnetic Reynolds number becomes (c.f.\ \citealt{TakeuchiShibata2001})}
\begin{equation}
R_{\rm m}=\frac{H_p c_s}{ \eta_O},
\end{equation}
{where $c_{\rm s}$ is the speed of sound. Indeed in these atmospheric layers the plasma $\beta$ may be larger than unity. This is especially important near the boundary of an isolated flux tube, where the Alfv\'en velocity is comparable to the sound speed. }We can estimate the pressure scale height in the solar atmosphere using the C7 model in \cite{AL2008}. Once the Ohmic magnetic resistivity $\eta_O$ has been calculated (see e.g.\ \citealt{KovityaCram1983, SinghKrishan:2010}) we can, for example, evaluate $R_{\rm mc}\sim 10^{\rm 5}-10^{\rm 4}$ in the solar photosphere at the solar temperature minimum \citep{KovityaCram1983}.

{To compare with the usual definition of the Lundquist number in Eq.~\eqref{standardLundquist}, we parameterize 
$$L v_A = H_p c_s L/(H_p\beta^{1/2})\simeq 10 H_p c_s.$$}
We then conclude that in order to discuss the onset of fast magnetic energy conversion through reconnection, we should consider how the choice of the parameters affects the critical threshold for the onset of the tearing instability, ultimately leading to the reconnection process.

\subsection{Magnetic Reynolds number in protoplanetary disk atmospheres.}
\label{LundquistDisks}
{In protoplanetary disks the magnetic fields' evolution depends on dynamical processes that may include
magneto-rotational instability (MRI) leading to channel flows and turbulence \citep{1991ApJ...376..214B,1991ApJ...376..223H,2005AIPC..784..475G,2000ApJ...534..398M,2001ApJ...561L.179S},
magnetically driven winds \citep{2009ApJ...691L..49S,1982MNRAS.199..883B,1995A&A...295..807F,2015ApJ...801...84G, 2020ApJ...896..126G,2024arXiv240103733I},
photoevaporative winds \citep{2021MNRAS.508.1675E,2023IAUS..362..294K},
vertical shear instability \citep{2013MNRAS.435.2610N,2016A&A...586A..33U,2018MNRAS.474.3110L,2020A&A...635A.190S,2023MNRAS.525..123H},
magnetic buoyancy and Parker instability \citep{2000ApJ...534..398M,2011ApJ...732L..30H}, and
magnetic reconnection \citep{2000ApJ...534..398M,2005ApJ...628L.155I,2013ApJ...767L...2M}.
Depending on the physical and chemical state of the environment, each of these processes can be of interest for local as well as global disk evolution, accretion of disk material onto the star, and the dispersal of the disk back into interstellar space.} {The dimensionless Reynolds and Elsasser numbers define how the fields evolve on large scales, but are insufficient for understanding how magnetic reconnection proceeds once a CS forms.  In this paper we focus on reconnection, so the magnetic Reynolds number defined above is a key parameter governing the CS thickness and so the role of the other non ideal effects.}\\

Most of the definitions of magnetic Reynolds numbers in protoplanetary disks have been provided in the context of the MRI and its saturation in MHD simulations (see the discussion in \citealt{2021ApJ...907...13P}). \citet{1998ApJ...506L..57S} defined $R_m=v^2_A/\Omega\eta_O=1$, where $\Omega$ is the angular velocity of the disk, below which MRI driven turbulence relaxes in channel flows. The latter define a new macroscopic lengthscale in the system, which can be disrupted by magnetic reconnection events. In the case of a poloidal field with zero vertical net flux, \citet{2000ApJ...530..464F} argued that the MRI can be sustained when the effective magnetic Reynolds number $R_m=c^2_s/\Omega\eta_O\le 10^4$. \citet{2017PhRvE..95c3202N} defined $R_m=L^2\Omega/\eta_O>10^3$, where $L$ is the size of the simulation domain in the radial direction, above which the magnetic turbulence induced by the MRI can be sustained, and energy is stored in the turbulent magnetic field, defining a new length at the eddy scale.
In general the definition of a magnetic Reynolds/Lundquist number depends on the dynamics and, in a protoplanetary disk, its calculation is difficult because of uncertainties on magnetic field detection \citep{2016ApJ...832...18L,2019A&A...624L...7V}. Recent measurements using the Zeeman effect provide estimates on the upper limits of magnetic fields of $|B|\sim 10^{-2} \rm{G}$ \citep{2019A&A...624L...7V}, significantly smaller than the values we see in the solar atmosphere. In this work we will vary the magnetic field in a range $B=10^{-4}-10^2 \rm{G}$ to cover a sufficiently large parameter space. The disk atmosphere is vertically stratified so that a scale height can be defined as $H_p:=c_s/\Omega$ where $c_s$ is the sound speed and $\Omega$ the orbital frequency.
The rotational speed of the atmospheric layer provides a torque, storing energy in the magnetic field lines, so we can think about the rotational speed as the main driver for the so-called accumulation or ``build-up'' phase (for magnetic field concentration and amplification in PPDs see \citealt{1981ARA&A..19..137P, 2007Ap&SS.311...35W, 2021JPlPh..87a2001P}).
Reconnection is particularly important in the regions where the plasma parameter $\beta\le1$, since most of the energy is stored in the magnetic field, which dominates the dynamics.
We will then adopt the Alfv\'en speed as the main driver velocity for magnetic field convection in the regions of interest, so the Lundquist number for the disk will be estimated here as $S=H_p v_A/\eta_O$. The latter will be calculated following \citet{2007Ap&SS.311...35W}, see the discussion on the model for PPDs adopted in this work, Sec.~\ref{Sect:Disks}.

\section{Partially ionized reconnection in the solar atmosphere}
\label{Sect:Sun}
In this section, we will discuss the regimes in which fast magnetic reconnection can be triggered in the solar photosphere and chromosphere.

\begin{table*}
  \centering
\pgfplotstableset{
    columns/0/.style={column name={{\bf Height}}},
    columns/1/.style={column name={{\bf $T$[K]}}},
    columns/2/.style={column name={{\bf $n_e$}}},
    columns/3/.style={column name={{\bf $n_n$}}},
    columns/4/.style={column name={{\bf $n_{HeI}$}}},
    columns/5/.style={column name={{\bf $n_{HeII}$}}},
    columns/6/.style={column name={{\bf $\nu_{ei}$}}},
    columns/7/.style={column name={{\bf $\nu_{en}$}}},
    columns/8/.style={column name={{\bf $\nu_{in}$}}},
    columns/dof/.style={int detect,column type=r,column name=\textsc{Dof}},
    columns/error1/.style={
        sci,sci zerofill,sci sep align,precision=1,sci superscript,
        column name=$e_1$,
    },
    columns/error2/.style={
        sci,sci zerofill,sci sep align,precision=2,sci 10e,
        column name=$e_2$,
    },
    columns/{grad(log(dof),log(error2))}/.style={
        string replace={0}{},
        column name={$\nabla e_2$},
        dec sep align,
    },
    columns/{quot(error1)}/.style={
        string replace={0}{},
        column name={$\frac{e_1^{(n)}}{e_1^{(n-1)}}$}
    },
    empty cells with={--},
    every head row/.style={before row=\toprule,after row=\midrule},
    every last row/.style={after row=\bottomrule}
}
\begin{filecontents*}{Pip-table-python-sun-I-withHeI.txt}
0.00000E+00   6.58300E+03   8.39700E+13   1.18700E+17   1.18000E+16   1.17000E+01   6.40000E+09   7.97000E+09   2.63000E+09
2.00000E+01   6.23100E+03   4.25800E+13   1.06000E+17   1.18000E+16   1.17000E+01   3.53000E+09   7.23000E+09   2.39000E+09
5.00000E+01   5.82600E+03   1.90900E+13   9.68000E+16   9.52000E+15   1.65000E-01   1.77000E+09   6.12000E+09   2.02000E+09
1.00000E+02   5.43100E+03   8.74800E+12   7.24400E+16   7.04000E+15   1.02000E-02   9.15000E+08   4.42000E+09   1.46000E+09
1.25000E+02   5.28800E+03   6.69600E+12   6.15000E+16   7.04000E+15   1.02000E-02   7.28000E+08   3.71000E+09   1.22000E+09
2.00000E+02   5.01000E+03   3.30800E+12   3.54900E+16   3.44000E+15   2.56000E-04   4.00000E+08   2.08000E+09   6.87000E+08
3.50000E+02   4.70000E+03   9.33700E+11   1.03400E+16   1.55000E+15   1.52000E-05   1.25000E+08   5.87000E+08   1.94000E+08
4.50000E+02   4.48500E+03   3.87100E+11   4.32200E+15   4.29000E+14   4.12000E-01   5.70000E+07   2.40000E+08   4.00000E+07
4.90000E+02   4.43500E+03   2.71800E+11   2.99300E+15   2.74000E+14   2.98000E+00   4.14000E+07   1.65000E+08   2.72000E+07
5.25000E+02   4.41000E+03   1.99600E+11   2.15500E+15   2.18000E+14   5.78000E+00   3.12000E+07   1.19000E+08   1.96000E+07
5.60000E+02   4.40000E+03   1.46700E+11   1.54500E+15   1.73000E+14   9.62000E+00   2.44000E+07   8.50000E+07   1.40000E+07
6.15000E+02   4.43500E+03   9.16300E+10   9.05800E+14   1.09000E+14   1.89000E+01   1.44000E+07   5.00000E+07   8.24000E+06
6.60000E+02   4.51000E+03   6.42000E+10   5.82500E+14   6.94000E+13   3.21000E+01   9.83000E+06   3.24000E+07   5.34000E+06
7.50000E+02   4.84000E+03   4.76200E+10   2.40000E+14   2.85000E+13   6.09000E+01   7.00000E+06   1.40000E+07   2.30000E+06
8.00000E+02   5.09000E+03   5.45800E+10   1.49500E+14   1.91000E+13   6.87000E+01   7.11000E+06   8.84000E+06   1.45000E+06
8.54000E+02   5.43000E+03   8.05500E+10   9.08500E+13   1.30000E+13   8.00000E+01   9.50000E+06   5.55000E+06   9.15000E+05
9.00000E+02   5.72000E+03   1.16200E+11   6.08600E+13   8.97000E+12   9.38000E+01   1.30000E+07   3.80000E+06   6.29000E+05
9.46000E+02   5.96900E+03   1.51300E+11   4.17800E+13   5.26000E+12   1.20000E+02   1.54000E+07   2.67000E+06   4.40000E+05
9.71000E+02   6.10000E+03   1.72700E+11   3.42800E+13   5.26000E+12   1.20000E+02   1.70000E+07   2.22000E+06   3.70000E+05
1.06500E+03   6.40000E+03   2.09100E+11   1.72200E+13   2.68000E+12   1.69000E+02   1.91000E+07   1.14000E+06   1.89000E+05
1.14300E+03   6.53100E+03   1.99500E+11   1.00900E+13   1.40000E+12   6.86000E+02   1.76000E+07   6.80000E+05   1.11000E+05
1.39800E+03   6.61000E+03   1.43800E+11   1.88000E+12   4.22000E+11   5.63000E+02   1.25000E+07   1.30000E+05   2.10000E+04
1.52000E+03   6.62300E+03   1.42300E+11   7.95600E+11   1.35000E+11   9.59000E+02   1.25000E+07   5.34000E+04   8.85000E+03
1.72200E+03   6.64300E+03   1.02700E+11   1.91600E+11   6.66000E+10   2.89000E+03   8.95000E+06   1.30000E+04   2.13000E+03
1.94600E+03   6.66700E+03   5.52600E+10   4.37100E+10   1.56000E+10   1.35000E+04   4.82000E+06   3.00000E+03   4.87000E+02
2.09800E+03   6.70000E+03   3.29300E+10   2.25800E+10   2.53000E+03   6.16000E+00   2.90000E+06   1.52000E+03   2.52000E+02
\end{filecontents*}
\pgfplotstabletypeset[
    1000 sep={\ },
    columns/info/.style={
        fixed,fixed zerofill,precision=1,showpos,
        column type=r,
    }]{Pip-table-python-sun-I-withHeI.txt}
        \caption{Input parameters for the model vs.\ height in the solar atmosphere.  The height has units of $10^5$~cm, the temperature $T$ is in Kelvin, the number densities are in cm$^{-3}$. $n_{HeI}$ is the neutral helium abundance while $n_{HeII}$ is the ionized helium abundance. The electron-ion ($\nu_{ei}$), electron-neutral ($\nu_{en}$), and ion-neutral ($\nu_{in}$) collision frequencies are in $s^{-1}$, see the text for how these quantities are estimated.}
        \label{table1}
\end{table*}

\begin{table*}
  \centering
\pgfplotstableset{
    columns/0/.style={column name={{\bf Height}}},
    columns/1/.style={column name={{\bf $\eta_O$}}},
    columns/2/.style={column name={{\bf $\eta_H$}}},
    columns/3/.style={column name={{\bf $\eta_{AD}$}}},
    columns/4/.style={column name={{\bf $v_{A}$}}},
    columns/5/.style={column name={{\bf $H_p$}}},
    columns/6/.style={column name={{\bf $S$}}},
    columns/7/.style={column name={{\bf $R_m$}}},
    columns/dof/.style={int detect,column type=r,column name=\textsc{Dof}},
    columns/error1/.style={
        sci,sci zerofill,sci sep align,precision=1,sci superscript,
        column name=$e_1$,
    },
    columns/error2/.style={
        sci,sci zerofill,sci sep align,precision=2,sci 10e,
        column name=$e_2$,
    },
    columns/{grad(log(dof),log(error2))}/.style={
        string replace={0}{},
        column name={$\nabla e_2$},
        dec sep align,
    },
    columns/{quot(error1)}/.style={
        string replace={0}{},
        column name={$\frac{e_1^{(n)}}{e_1^{(n-1)}}$}
    },
    empty cells with={--},
    every head row/.style={before row=\toprule,after row=\midrule},
    every last row/.style={after row=\bottomrule}
}
\begin{filecontents*}{Pip-table-python-sun-II.txt}
0.00000E+00   4.83000E+07   7.10293E+07   3.08510E+05   6.66945E+05   4.00000E+07   1.10000E+06   1.11801E+06
2.00000E+01   7.00000E+07   1.37122E+08   6.52830E+05   6.75710E+05   3.78000E+07   7.20000E+05   7.07400E+05
5.00000E+01   1.17000E+08   2.93845E+08   1.56833E+06   6.88051E+05   3.53000E+07   3.90000E+05   3.83171E+05
1.00000E+02   1.72000E+08   5.87809E+08   3.99411E+06   7.18940E+05   3.30000E+07   2.34000E+05   2.35988E+05
1.25000E+02   1.90000E+08   7.42219E+08   5.79822E+06   7.32515E+05   3.21000E+07   2.10000E+05   2.04426E+05
2.00000E+02   2.11000E+08   1.25491E+09   1.46950E+07   8.00233E+05   3.04000E+07   1.70000E+05   1.70009E+05
3.50000E+02   2.15500E+08   3.07149E+09   8.78689E+07   9.46348E+05   2.85000E+07   1.50000E+05   1.50766E+05
4.50000E+02   2.17000E+08   5.70233E+09   6.00250E+08   1.09656E+06   2.72000E+07   1.40000E+05   1.40387E+05
4.90000E+02   2.15000E+08   7.27804E+09   1.00856E+09   1.16180E+06   2.70000E+07   1.40000E+05   1.39395E+05
5.25000E+02   2.11000E+08   8.97439E+09   1.57088E+09   1.21642E+06   2.70000E+07   1.39000E+05   1.40758E+05
5.60000E+02   2.10000E+08   1.10451E+10   2.43609E+09   1.27401E+06   2.70000E+07   1.42000E+05   1.41429E+05
6.15000E+02   2.00000E+08   1.50741E+10   4.85437E+09   1.37643E+06   2.70000E+07   1.51000E+05   1.49850E+05
6.60000E+02   1.90000E+08   1.88902E+10   8.16496E+09   1.47124E+06   2.74000E+07   1.65000E+05   1.61516E+05
7.50000E+02   1.21000E+08   1.95179E+10   1.50367E+10   1.68069E+06   3.00000E+07   2.80000E+05   2.87603E+05
8.00000E+02   8.26000E+07   1.47524E+10   1.54602E+10   1.79896E+06   3.08800E+07   4.44000E+05   4.48620E+05
8.54000E+02   5.28000E+07   8.60157E+09   1.24178E+10   1.92875E+06   3.30000E+07   7.70000E+05   7.62500E+05
9.00000E+02   4.00000E+07   5.29108E+09   9.46145E+09   2.04476E+06   3.47000E+07   1.10000E+06   1.08438E+06
9.46000E+02   3.38000E+07   3.63129E+09   8.73636E+09   2.17681E+06   3.62000E+07   1.40000E+06   1.37089E+06
9.71000E+02   3.14000E+07   2.99886E+09   8.03137E+09   2.22211E+06   3.70000E+07   1.53000E+06   1.53185E+06
1.06500E+03   2.73000E+07   2.01925E+09   8.69206E+09   2.45630E+06   3.88000E+07   1.89000E+06   1.89026E+06
1.14300E+03   2.51000E+07   1.80674E+09   1.09482E+10   2.66780E+06   3.96000E+07   2.12000E+06   2.11410E+06
1.39800E+03   2.50000E+07   1.53878E+09   2.87646E+10   3.29687E+06   4.01000E+07   2.21000E+06   2.16540E+06
1.52000E+03   2.50000E+07   1.23471E+09   3.96751E+10   3.68273E+06   4.02000E+07   2.20000E+06   2.17080E+06
1.72200E+03   2.50000E+07   1.20990E+09   9.87850E+10   4.10110E+06   4.03000E+07   2.20000E+06   2.19232E+06
1.94600E+03   2.50000E+07   1.61898E+09   2.50579E+11   4.52115E+06   4.04000E+07   2.23000E+06   2.19776E+06
2.09800E+03   2.50000E+07   2.28364E+09   5.18519E+11   4.50833E+06   4.07000E+07   2.23000E+06   2.27920E+06
\end{filecontents*}
\pgfplotstabletypeset[
    1000 sep={\ },
    columns/info/.style={
        fixed,fixed zerofill,precision=1,showpos,
        column type=r,
    }]{Pip-table-python-sun-II.txt}
        \caption{Derived physical parameters for the solar atmosphere, in c.g.s.\ units.
        The order of magnitude for the calculation of the non ideal Hall and AD coefficients is the same for both the values of magnetic fields $B_0$ at the bottom of the photosphere and so we chose $B_0=1200G$. The Alfv\'en speed in the table is also obtained using $B_0=1200G$ and the total (neutral and ion) mass density. The scale height $H_p$ is defined by Eq.~\eqref{accapiSUN}. The Lundquist number $S$ is calculated using $v_A$, $H_p$, and $\eta_O$ while the Reynolds number $R_m$ is calculated using the speed of sound $c_s$ in place of the Alfv\'en speed.}
        \label{table2}
        \end{table*}

\subsection{Lower solar atmospheric structure.}
\begin{figure}
 \includegraphics[width=85mm]{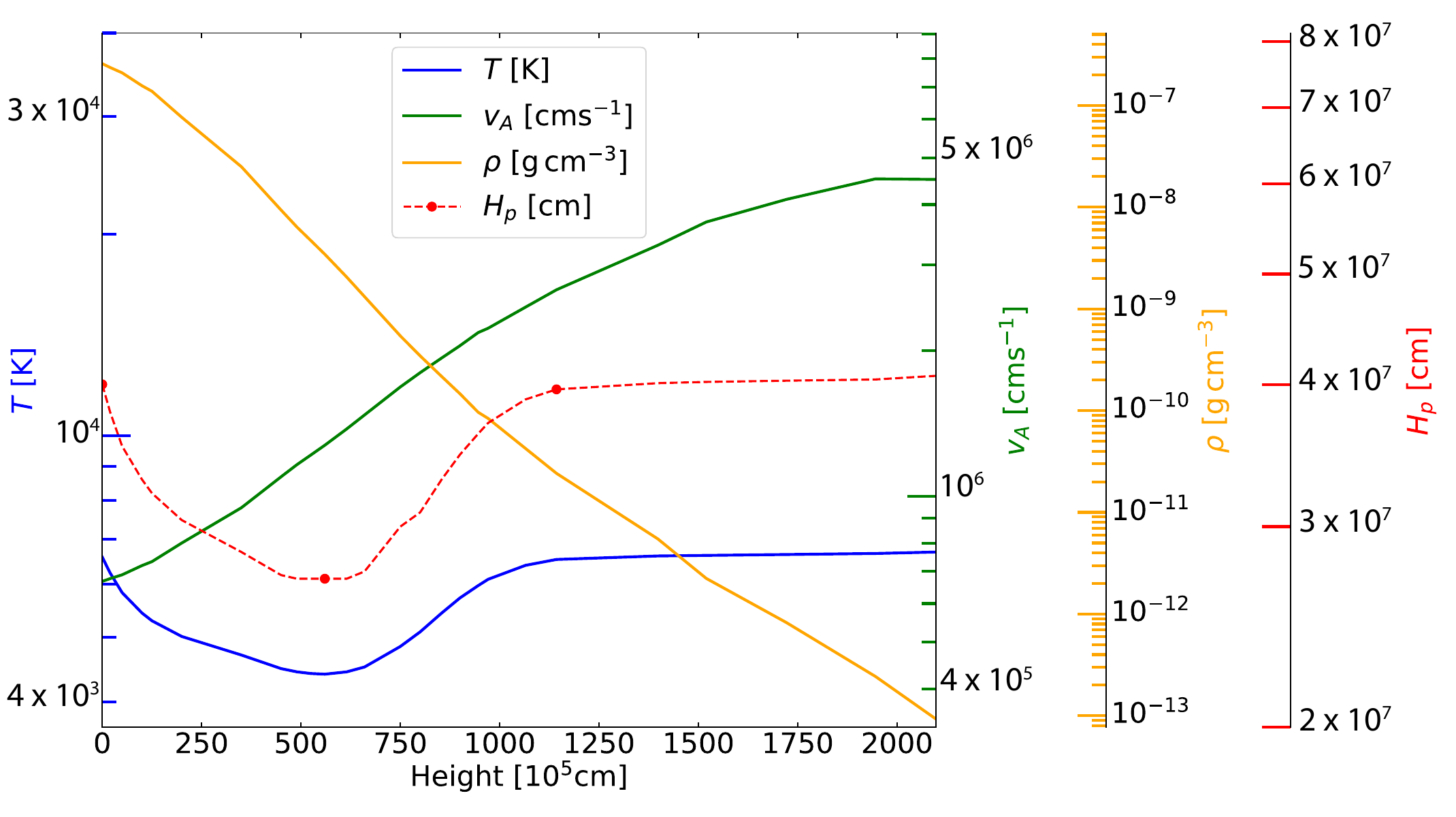}
 \caption{Temperature $T$, Alfv\'en speed $v_A$, total density $\rho$, and pressure scale height $H_p$, all in c.g.s.\ units, as functions of height in the solar atmosphere using Tab.~1.}
 \label{1DSUN}
\end{figure}

The model we adopt for the solar atmosphere is the one labeled C7 in \citet{AL2008}, where the quantities are functions only of height in the solar atmosphere.  The number densities of ions and neutrals are listed in Tab.~\ref{table1} assuming quasi-neutrality $n_i\sim n_e$.
As the most abundant species in the solar atmosphere are hydrogen and helium, the neutral and ionized helium abundances $n_{\rm HeI}$ and $n_{\rm HeII}$ are also provided.  The ionized helium abundance is orders of magnitude lower than the electron abundance (and so positive ion abundance, under quasi-neutrality), though helium contributes noticeably to the neutral population.  Henceforth we will adopt hydrogen as the main neutral species. 
The interspecies collision frequencies $\nu_{ei}$, $\nu_{en}$, $\nu_{in}$ are calculated using \citet{Zhdanov1962}, for which the collision cross-sections are obtained from \citet{VranjesKristic2013}. 
In Tab.~\ref{table1} we also list the temperature at each height.  The temperature profile plotted in Fig.~\ref{1DSUN} shows a minimum about 500~km above the solar surface.

Magnetic field at a given height is evaluated using the relation \citep{2005A&A...442.1091L}
\begin{equation}
B=B_s\left(\frac{\rho}{\rho_s}\right)^{\alpha},
\end{equation} 
where $B_s$ is the magnetic field at the solar surface, defined as the surface with optical depth equal to unity (bottom of the photosphere). We choose $\alpha = 0.3$ so the magnetic field weakens with height.  We experiment with $\alpha$ over the range $0.3$ to $0.6$ \citep{2005A&A...442.1091L} with only minor impacts on the results.  We consider two different field strengths at the solar surface, $B_s=1200$~G and $B_s=2200$~G \citep{2010NewA...15..119S}.

In Fig.~\ref{1DSUN} we show also the total density, from which we derive the Alfv\'en speed profile.  The pressure scale height is
\begin{equation}
\label{accapiSUN}
H_p=\frac{\mathcal{R} T}{\mu g},
\end{equation} 
where $\mathcal{R}$ is the gas constant, $\mu$ the mean mass per mole of atmospheric plasma, and $g$ the local acceleration due to gravity (Sec.~\ref{ReynoldsSolar}). 
We now have all the quantities needed to calculate the critical scales in Eq.~\eqref{CriticalScalesPartial}.

\subsection{Critical scales for the onset of fast magnetic reconnection in the lower solar atmosphere.}
\label{ResultSun}
In Tab.~\ref{table2} we listed the resulting Ohmic diffusivity and, for comparison, the Hall and AD coefficients. The Hall coefficient is the same order of magnitude as the Ohmic magnetic resistivity. The AD depends on the magnetic field and increases with height in the atmosphere. 
While these quantities tell us which diffusion process is dominant, as discussed before, diffusion is slow compared to advection and fast reconnection.  We will come back to this later in this section.
Magnetic Reynolds numbers, necessary to calculate the critical CS for the IT mode, inherited the contribution due to the ion-neutral collisions from the diffusivity.

\begin{figure}
\includegraphics[width=95mm]{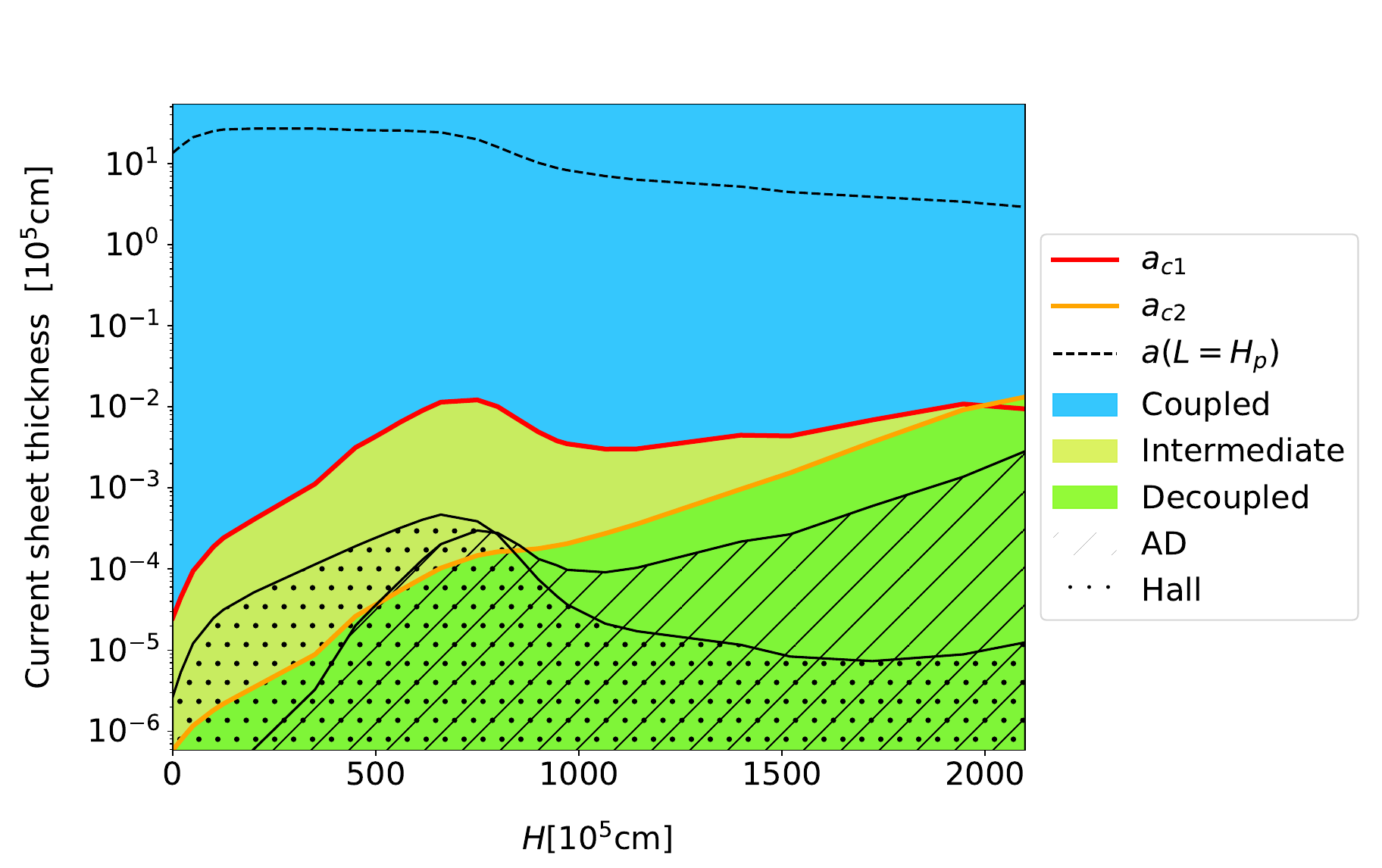}
\caption{Reconnection regime versus height in the solar atmosphere and thickness of the CS.  The critical CS thicknesses $a_{c1}$ and $a_{c2}$ are the red and orange lines, respectively.  Blue shading shows the regime where ions and neutrals are well-coupled by collisions, yellow the intermediate regime, and green the regime where ions and neutrals are decoupled and reconnection proceeds at the faster Alfv\'en speed of the ions alone.  The dotted overlay marks where the Hall effect is expected to play a role, and the barred overlay is the where AD must be considered.  The black dashed line near top indicates the thickness of a CS having length $L=H_p$ in the resistive coupled regime.  A CS of length $L$ lying above the dashed line can reconnect slowly, but once fast reconnection begins at the dashed line, thinner CSs below the dashed line can form only through recursive (secondary) reconnection events.}
\label{2Ddiagram-SUN}
\end{figure}

In Fig.~\ref{2Ddiagram-SUN} we show the results for the transition between different reconnection regimes (color coded), as a function of height in the solar atmosphere. The coupled regime (blue area) extends down to CSs with thickness of $a \ll 1$~km for all the heights, both in the photospheric as well as in the chromospheric layer. 
The Lundquist numbers and the magnetic Reynolds numbers resulted to be of the same order of magnitude, and so we used $S$ for the results shown in Fig.~\ref{2Ddiagram-SUN}.

The critical length scale $a/H_p\sim S^{-1/3}(\rho_n / \rho_i)^{1/6}$ associated with the longest available CS, i.e.\ limited by $L<H_p$ at each height, is shown in Fig.~\ref{2Ddiagram-SUN} (dashed black line). This means there is a range of CS thicknesses in the solar atmosphere $a\lesssim 100$~km for which reconnection begins with ion-neutral coupled dynamics.  When the magnetic field dominates the dynamics ($\beta<1$), larger CSs are likely to form.
{Scales down to $a\sim 100$~km could be resolved with DKIST or SOLAR-C \citep{10.1117/12.2055366,10.1117/12.2560887,2020SoPh..295..172R}, while smaller-scale features may be hidden since the photon mean free path $\sim 100$~km at the photosphere \citep{Judge2015}.}
As explained in Sec.~\ref{cascade-fractal}, we expect primary CSs to break into secondary CSs that are unstable to IT in the intermediate or decoupled regime (see also \citealt{Teneranietal:2015, Singhetal:2019}).

\subsection{{Hall and Ambipolar diffusion effects.}}

{In Fig.~\ref{2Ddiagram-SUN} the dotted and barred areas indicate where the Hall and AD contributions to magnetic reconnection might not be negligible.  For the AD to be important, a degree of coupling between neutrals and ions is necessary, so ambipolar effects are not relevant in the decoupled regime.  The upper edge of the barred area is calculated using Eq.~\eqref{ADimportant}.  
If and only if there is a neutral point, AD steepens the current profile in the forming current sheet (\citealt{2020RSPSA.47690867N, 1997ApJ...478..563Z,2023arXiv231214076T}), once the dynamics of reconnection are taken into account. 

The upper limit of the dotted area is defined  by the local ion inertial length $d_i$, for $H^+$ ions. The Hall effect on its own does not break the frozen-in condition, as field lines remain frozen to the electron flow, but the Hall term destabilizes reconnection and affects tearing and the IT instability criterion (\citealt{Terasawa:1983,Puccietal:2017, 2020ApJ...902..142S}). This means that in the dotted region in Fig.~\ref{2Ddiagram-SUN}, the critical length-scale at which reconnection becomes efficient will also depend on the ion inertial length, and the CS will become unstable at slightly smaller aspect ratio than the simple resistive case, with corrections provided in \cite{Puccietal:2017}. 
Still, even if both the AD and Hall diffusion are relevant to the reconnection process, the energy conversion will proceed at a pace determined by the regime in which the reconnection event is happening, i.e.\ coupled, intermediate or decoupled, see Fig.~\ref{RegTransition}.}

\subsection{Fractal reconnection scenario.}

As discussed in Sec.~\ref{cascade-fractal}, secondary reconnection is likely to occur in the intermediate or the decoupled regime in partially-ionized plasmas.  In the lower solar chromosphere we have from Tab.~\ref{table1} that $\rho/\rho_i \sim n_n/n_e={10^{3}}$ since the plasma is quasi-neutral and $m_i\sim m_n$.  The density ratio falls to $\rho/\rho_i \sim 1$ in the fully-ionized upper solar atmosphere.  The Lundquist number $S$ ranges from $S=10^5-10^6$ (Tab.~\ref{table2}).  Choosing $S=10^5$ and $\rho/\rho_i=10^3$, for Eq.~\eqref{secondaryTrigger2} to be satisfied $\tau_{A}\le 100 \tau_{ni}$.  In the coupled regime (blue region of Fig.~\ref{RegTransition}), as $a_c$ approaches $a_{c1}$ then $\gamma \sim \tau_{A}\sim \tau_{ni}$.  In the first equality here we use the fact that the CS has already developed into fast IT in the primary reconnection stage.  Then Eq.~\eqref{secondaryTrigger2} is satisfied and secondary reconnection begins in the intermediate regime.

That reconnection proceeds faster at smaller scales, where the ions and neutrals decouple, leads to runaway acceleration of the ions that could produce tails in the ion energy distribution like those needed for a significant contribution to coronal heating from velocity filtration \citep{Scudder92,ScudderOlbert79a,ScudderOlbert79b}.

\section{Reconnection in protoplanetary disk atmospheres.}
\label{Sect:Disks}

{Reconnection has not been clearly identified in PPDs.  Even the global angular scales are unresolved, since the nearest PPDs are more distant than the Sun by seven orders of magnitude.  In this section we discuss where reconnection is likely to occur in PPDs, how reconnection events can be initiated in the disk context, and how the resulting magnetic energy dissipation may impact the disk and its atmosphere.}

We apply the reconnection picture described above in sections~\ref{PIPSummary} and~\ref{cascade-fractal} to a model disk orbiting a star of $M = M_{\odot}$ that is about $10^6$~yr old, an age at which such disks are typically accreting onto their central stars.  We focus on conditions near 2.5~au from the star, since (1) the disks' central few~au is the launching point for winds observed through optical line emission \citep{2016ApJ...831..169S, 2019ApJ...870...76B,2023ASPC..534..567P} and (2) the disk atmosphere in this same distance range is the source of a multitude of mid-infrared molecular emission lines that carry information on the excitation conditions \citep{2011ApJ...731..130S, 2014prpl.conf..363P, 2015A&A...582A..88W}.

{At 2.5~au, the dipole component of the star's magnetic field is millions of times weaker than at the stellar photosphere and the higher-multipole components are weaker still.  Measured field strengths at young near-Solar-mass stars' photospheres are in the kilogauss range \citep{2011ApJ...729...83Y}, corresponding to fields at au distances less than a milligauss.  Comparing this to the disk fields swept in with the gas during the collapse of the parent molecular cloud core \citep{2016A&A...587A..32M} and the gauss-range fields inferred for the asteroid belt's location in the early Solar system \citep{2021SciA....7.5967W}, it seems likely that the stellar fields outside 1~au are negligible compared with the fields intrinsic to the disk.}

\subsection{{Turbulence as a driver for magnetic reconnection.}}
\label{MagRecConfDisks}
{
Though MRI is not thought to be active at the midplane of protoplanetary disks (\citealt{2015ApJ...801...84G,2019ApJ...872...98M,2024arXiv240103733I}), 
it can be triggered in any layer where both of the Elsasser numbers exceed unity and the magnetic pressure is less than the gas pressure (\citealt{2022arXiv220309821L}).
The nature of the magnetic dynamics and in particular the presence of MRI-driven turbulence depend on the heating and cooling processes and chemical reactions included, since these govern the non-ideal coefficients in the induction equation \citep{2015ApJ...801...84G,2014ApJ...791..137B, 2020ApJ...896..126G,2024arXiv240103733I}.
The static disk model we employ has more species and more heating and cooling processes than can typically be treated in dynamical calculations.  The coupling of chemistry to thermodynamics leads to a disk with a hot, extended, ionized atmosphere.  As a result, we shall see below that the criteria for MRI are met in a layer spanning heights above the midplane from 0.35 to 0.75~au.}

{If MRI is present some of the magnetic fields generated in the MRI turbulent region will rise into the atmosphere due to their buoyancy (\citealt{2000ApJ...534..398M}) and through the Parker instability (\citealt{2011ApJ...732L..30H}).
Reconnection can develop in turbulent environments driven by the MRI \citep{2013ApJ...767L...2M,2000ApJ...534..398M,2005ApJ...628L.155I}, and MRI-driven currents and reconnection events can cause enough ionization to sustain MRI turbulence itself \citep{2005ApJ...628L.155I}.  Recent work on MHD turbulence shows that eddies develop strong anisotropies in the inertial range and that magnetic reconnection in the ideal regime can even break the cascade leading to dissipation \citep{BoldyrevLoureiro20}.  The energy released potentially contributes to the heating of protostellar winds and jets \citep{2000ApJ...540..372K}.}

{The role and importance of reconnection may not be readily observable in disk simulations because limited spatial resolution leads to underestimating Lundquist and magnetic Reynolds numbers in certain regions, and also because time-averaging magnetic fields and assuming axisymmetry can hide the perturbations driving reconnection.}

\subsection{{Magnetic field tangling and CSs in the protoplanetary disk atmosphere.}}

The magnetic topology in protoplanetary disks is driven by the field's coupling to flows arising from differential rotation and turbulence.  CSs may form as natural boundaries separating flux volumes rooted in different disk regions. For example, if the disk atmosphere has a beta transition at which gas and field are coupled, then any flows in the gas will move atmospheric magnetic field footpoints. In a system where a form of line-tying exists, differential rotation of footpoints leads naturally to the formation of CSs separating field lines rooted at different radii, or in regions separating open and closed field lines in the magnetically linked star-disk system \citep{1995MNRAS.275..244L, 1996MNRAS.281..219T, 1998ApJ...500..703R,2006astro.ph..7656U,
2017ApJ...847...46T}.

{One large-scale CS is that formed by the hourglass magnetic configuration \citep{2014prpl.conf..173L} of the disk’s magneto-centrifugal wind when acted upon by the differential orbital rotation. This leads naturally to toroidal fields with opposite signs in the upper and lower hemispheres.  CSs can also form at the boundary between the so-called ``dead zone'' and the wind launching region, where magnetic field is wound up by differential rotation.  In diffusive MHD calculations including those by \citet{2015ApJ...801...84G, 2020ApJ...896..126G} and \citet{2024arXiv240103733I}, this CS is offset from the midplane and occurs in the atmosphere on either the upper or lower face of the disk.}

\subsection{{Disk chemistry and radiation.}}
\label{diskmodel}
{We assess reconnection in protoplanetary disks in the context of a thermochemical disk model \citep{2019ApJ...885..146R} having mass $0.01\mathrm{M}_{\odot}$ and orbiting a 1$\mathrm{M}_{\odot}$ star whose optical, UV, and X-ray spectrum are those of the well-studied young star TW~Hya.}

{As detailed in \citet{2019ApJ...885..146R}, we determine the disk's density structure by solving for vertical hydrostatic balance under a surface density constraint set by the disk mass. 
Gas temperature is found by balancing heating by dust collisions, UV, and X-rays with cooling by line and continuum emission from various ions, atoms, molecules, and grains, whose abundances are determined by solving a chemical network.  The network includes $\sim 6000$ gas-phase reactions, photo-reactions (UV and X-rays), cosmic ray ionization, and grain surface reactions among $\sim 800$ species.  The gas density, temperature, and chemistry, and the dust size distribution are all coupled.  The grain size distribution at each height is determined by balancing fragmentation, coagulation, and settling.  The cooling is computed using non-LTE radiative transfer in the main spectral lines treating collisional and radiative processes.  The dust's thermal radiation is also included.}

\begin{figure}
\hskip -0.7cm
 \includegraphics[width=98mm]{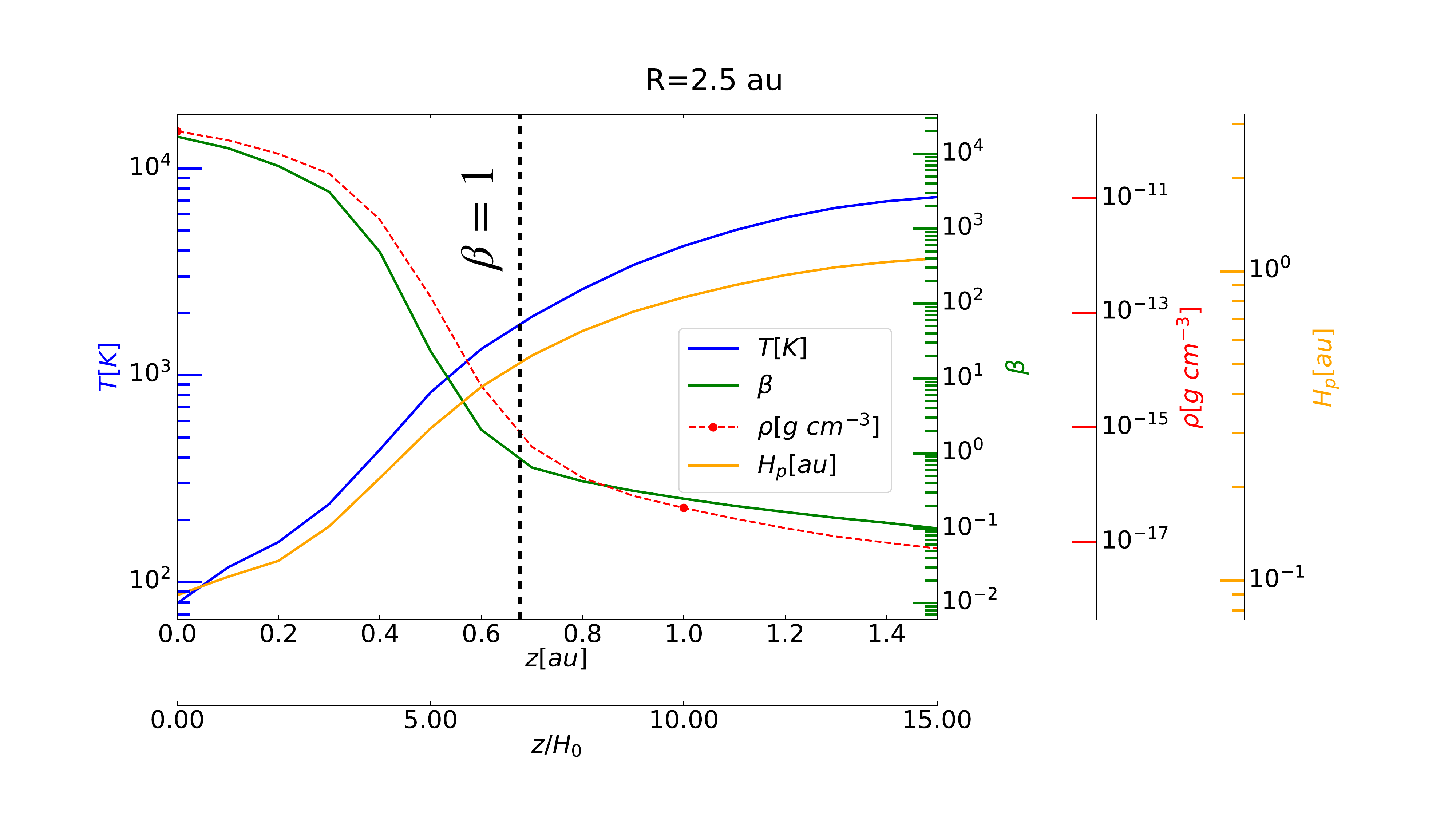}
 \caption{Conditions in the model protoplanetary disk versus height $z$ at distance 2.5~au from the Solar-mass star, obtained from the thermochemical calculation: temperature $T$ in Kelvin, ratio of gas to magnetic pressure $\beta$, total (neutral plus ion) mass density $\rho$, and scale height $H_p=c_s/\Omega$ where $\Omega$ is the orbital frequency.  {The black dashed vertical line marks where $\beta=1$}. {The second horizontal axis shows the height in units of the midplane pressure scale height $H_0=H_p(z=0)$.  Since the scale height increases with $z$, the domain spans a large multiple of the midplane scale height.}}
 \label{nonIDEAL-Disk}
\end{figure}

\begin{figure}
 \includegraphics[width=85mm]{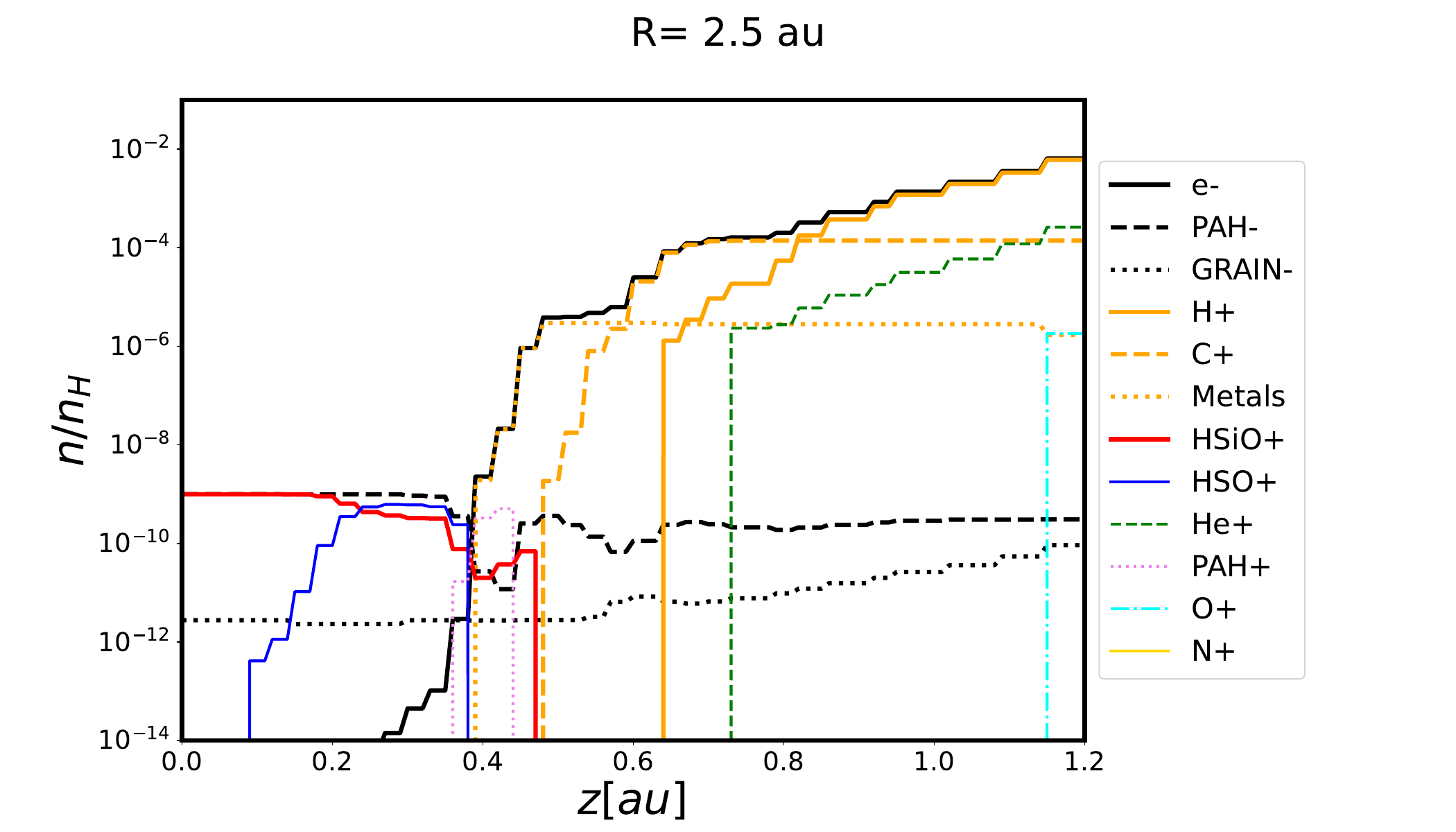}
 \caption{Main charge carriers' abundances versus height in the model protoplanetary disk at a distance 2.5~au from the Solar-mass star.  The plot runs from the midplane {up to $z=1.2$~au, above which the abundances are nearly uniform due to the high ionization rate}.  All abundances are relative to the number density of hydrogen nuclei, $n_H$.  ``Metals'' is the sum of Mg$^+$ and Fe$^+$ ions.\\}
 \label{Abundance}
\end{figure}

{The density structure, heating and cooling, and chemistry are computed iteratively to mutual consistency.  Fig.~\ref{nonIDEAL-Disk} shows the resulting variation with height of the total (ions plus neutrals) density and temperature. 
Also plotted there is the pressure scale height $H_p$, used below as an estimate of the largest coherent scale for CS formation $L=H_p$ as discussed in Sec.~\ref{LundquistDisks}.  The main charge carriers' abundances versus height are shown in Fig.~\ref{Abundance}.}

The model disk extends from the midplane up to heights above the base of the layer where the star's X-ray and FUV photons can heat the gas to near $10^4$~K 
while it remains neutral. The temperature at these heights is too low for the hot gas to readily escape, since the virial temperature at our fiducial distance of 2.5~au from the star is $\gtrsim 20,000$~K \citep{Gorti2015, 2023ASPC..534..567P}. Photoevaporative escape from our fiducial radius is possible at larger heights where the gas is fully ionized, but for the range of heights $z \lesssim 1$~au we are concerned with here, the disk structure does not significantly depart from hydrostatic equilibrium.
The thermal structure implies the base of the photoevaporative wind lies at a height 1.14~au and the launching speed is just 16\% of the thermal speed, by the methods in \citet{2004ApJ...611..360A} and \citet{Gorti2015}.  This slow outflow will have only slight effects on the disk structure in the region of interest here (we focus on $z \sim 0.6$\ au where $\beta=1$).  Examining how reconnection operates in this hydrostatic model is thus sufficient.

\subsection{{Modeling dust grains.}}
\label{modelinggrains}
{Dust grains' surfaces catalyze chemical reactions including the recombination of free charges.  Dust also enables photoelectric heating and continuum radiation cooling.  Dust thus can have major effects on the induction equation's non-ideal coefficients \citep{2007Ap&SS.311...35W, 2022ApJ...934...88T}.  Our model disk has a ratio of dust to gas surface density $\Sigma_{\rm dust}/\Sigma_{\rm gas}=0.01$, near the value in the interstellar medium.  The silicate and carbonaceous grains' size distribution when integrated through the disk thickness corresponds to equilibrium between collisional aggregation and fragmentation, with number density $n_d(a)\propto a^{-3.5}$ over the range $0.005~\mu{\rm m}<a<1$~cm.  Each grain size's vertical distribution within the gas column is determined by balancing settling with stirring by weak turbulence, so that the bigger grains are concentrated near the midplane as observed \citep{2008ApJ...683..479B, 2008A&A...489..633P}.
The grain temperature is size- and composition-dependent and the solids are released into the vapor phase from grains whose temperature exceeds the sublimation threshold of $\sim 1500$K.}

\subsection{{Non-ideal coefficients and Elsasser numbers.}}
\label{ElsasserNumbers}
{Both the AD and the reconnection heating rate depend on the magnetic field strength.  We estimate the field in our model disk annulus as follows.  Protoplanetary disks deliver material to the surfaces of their central young stars at rates typically around ${\dot M}=10^{-8}$ Solar masses per year at age one million years \citep{2020ApJ...893...56M}.  If this accretion flow is driven by the torque from a magneto-centrifugal wind, the magnetic field threading the disk cannot be weaker than $B > (2{\dot M}\Omega/{\sqrt{3}R})^{1/2}$ (\citealt{2007Ap&SS.311...35W} and \citealt{2009ApJ...701..737B} Eq.~7).  For the disk annulus located $R=2.5$~au from the Solar-mass star, this amounts to a lower limit on the field strength $B > 30$~milligauss.}

{While measurements of magnetization in primitive meteorites suggest fields up to about 500~milligauss at 1 to 3~au from the young Sun in the solar system's first million years \citep{2021SciA....7.5967W}, we give our model disk annulus a field with the minimum strength of 30~milligauss to obtain a lower bound on the reconnection heating rate.}

{Empirical evidence on protoplanetary disks' magnetic field geometries is so limited that a wide range of possibilities remains open.  We therefore simply assume the total field is uniform in height to make exploration of the parameter space straightforward.  Fig.~\ref{nonIDEAL-Disk} shows that the 30-milligauss field yields a ratio of gas to magnetic pressure $\beta \sim 10^4$ at the midplane.  This ratio falls with height, being less than unity above $z=0.6$~au.  The two other magnetic field properties needed below are the magnitude of the vertical field component, which governs whether magneto-rotational turbulence is present, and the magnitude of whichever field component is involved in reconnection.  Again seeking the simplest reasonable choice since the evidence is too limited to support greater complexity, we assume both the field's vertical component and its reconnecting component are one-third of the total field, or 10~milligauss.}

{We can now determine the coefficients in the induction equation's non-ideal terms at each height in the model disk annulus using Eqs.~(25)--(31) in \cite{2007Ap&SS.311...35W}.  Included are the contributions to the electric current from all the charged gas-phase and grain species tracked in the thermochemistry code; inter-species collision frequencies from \cite{1999MNRAS.303..239W}; and the magnetic field described above.}

\begin{figure}
\includegraphics[width=80mm]{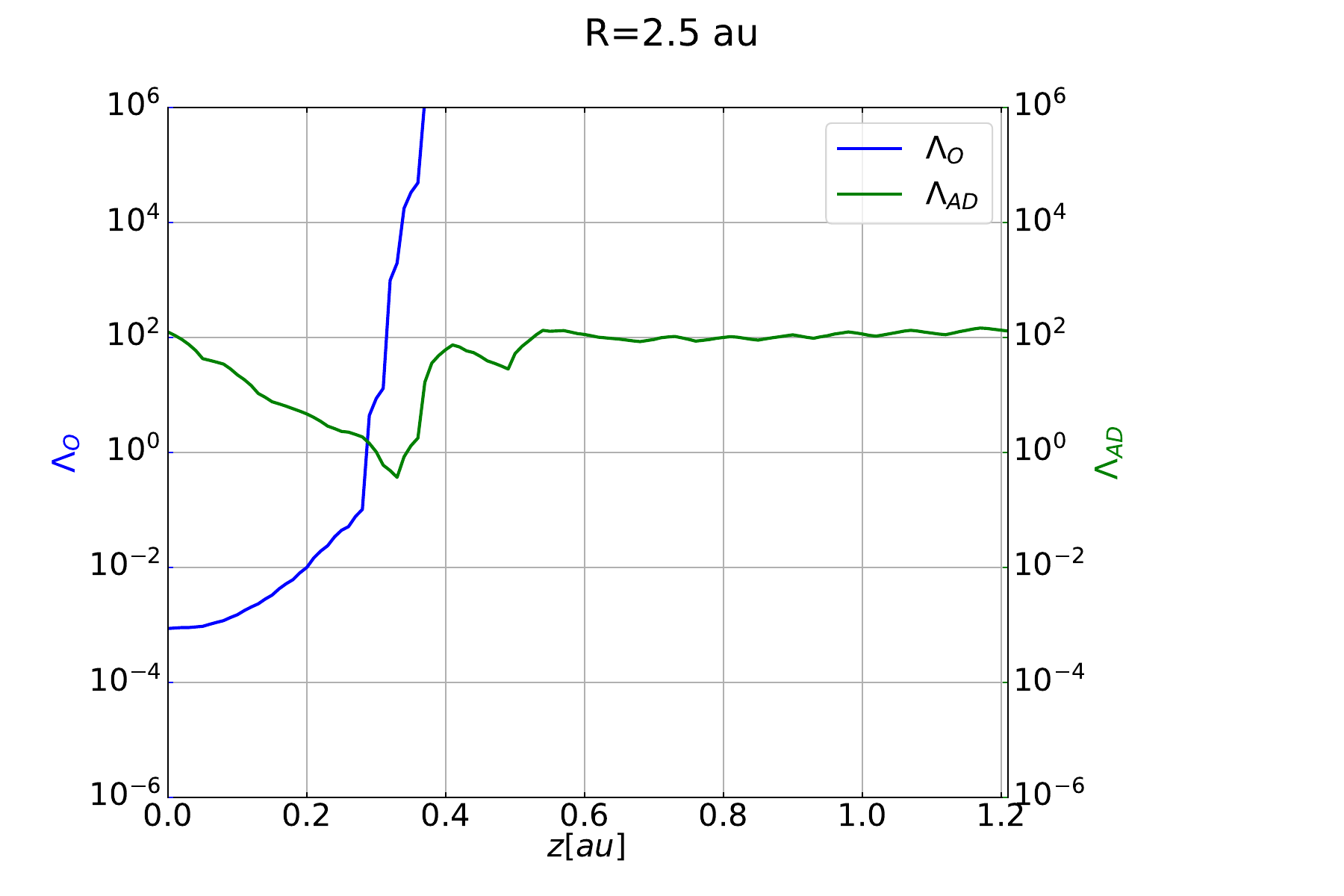}
\caption{{Dimensionless Ohmic and ambipolar Elsasser numbers versus height in the model protoplanetary disk at a distance 2.5~au from the Solar-mass star.  A necessary condition for magneto-rotational instability leading to turbulence is that both Elsasser numbers exceed unity.  This condition is met above 0.35~au.  The plot extends from the midplane to $z=1.2$~au, above which the Elsasser numbers are almost independent of height.}}
\label{ElsasserNum}
\end{figure}

{The resulting non-ideal coefficients yield the Ohmic and ambipolar Elsasser numbers $\Lambda_{O/AD}=v^2_{Az}/(\Omega \eta_{O/AD})$ that govern the magnetic field dynamics away from current sheets in the disk context \citep{2022arXiv220309821L}.  These two dimensionless numbers are plotted versus height in Fig.~\ref{ElsasserNum}.  The ambipolar Elsasser number is the smaller of the two and less than unity, hence the limiting factor for MRI, only in a thin layer around height 0.3~au.  Above about 0.35~au both Elsasser numbers exceed unity, a necessary condition for MRI to operate.  A further requirement is plasma $\beta$ greater than unity, which holds below 0.66~au.  The layer in the model disk atmosphere between 0.35 and 0.66~au thus meets the conditions for the onset of turbulence driven by the MRI.} 
We have checked that these results depend weakly on the choice of field strength.  For example, if the field were twice the minimum capable of driving the typical accretion flow, the two Elsasser numbers would both exceed unity above 0.34~au and the plasma beta exceed unity below 0.58~au, so the layer of the disk atmosphere in between would remain subject to MRI.  The stronger field would enable reconnection heating rates several times greater than we discuss below.

\subsection{{Critical CS thickness and regimes for the onset of fast reconnection.}}
\label{DiskCT}

\begin{figure}
 \includegraphics[width=90mm]{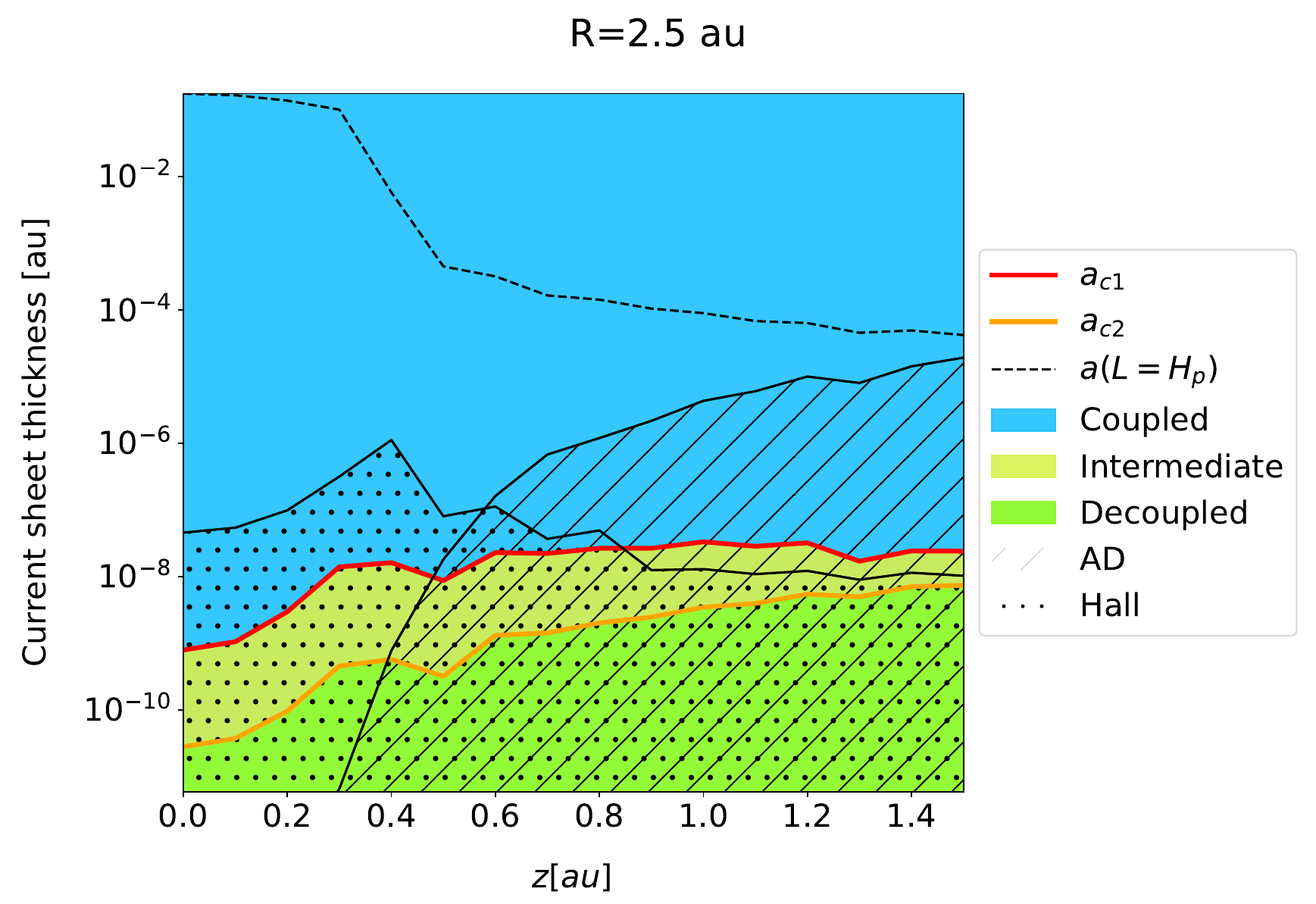}
 \caption{Reconnection regime versus height in the model protoplanetary disk and thickness of the CS.  The distance from the central star $R=2.5$~au.  The fully-coupled, intermediate, and decoupled regimes are shown in blue, yellow, and green, respectively.  A black dashed line marks the greatest CS thickness $a$ available in the system given the maximum coherent length scale $L=H_p$, calculated as $a\sim H_p S^{-1/3}(\rho_n/\rho_i)^{1/6}$ (Fig.~\ref{RegTransition}).  The dotted and barred overlays indicate where the Hall effect and AD, respectively, are expected to play roles in determining the critical threshold for fast magnetic reconnection.}
 \label{Disk-Crit-Length}
\end{figure}

Next we discuss the dominant effects in reconnection dynamics, which are determined not by the Elsasser numbers but by the Alfv\'en speed and diffusion coefficients as described in Sec.~\ref{PIPSummary}.  We calculate for the model protoplanetary disk the critical CS thicknesses $a_{c1}$ and $a_{c2}$ at which reconnection transitions from the fully-coupled down to the intermediate regime and from the intermediate down to the decoupled regime following Fig.~\ref{RegTransition}.  These two scales vary with height $z$ in the disk atmosphere as shown in Fig.~\ref{Disk-Crit-Length} by the red and the orange line, respectively.  They divide the fully-coupled regime in blue from the intermediate regime in yellow and the decoupled regime in green.  

A black dashed line near the top of Fig.~\ref{Disk-Crit-Length} shows the thickness of the longest CS available at each height in the disk.  This CS has length $L\sim H_p$ and thickness given by the coupled regime critical reconnection scale $a_c$ in the blue row of Fig.~\ref{RegTransition}.  The black dashed line remains in the blue zone even high in the disk atmosphere above 1~au, where the stellar X-rays make ionized hydrogen the main charge carrier (Fig.~\ref{Abundance}).  Thus even here, magnetic reconnection can be triggered in the decoupled regime on the longest CSs.

The scale height $H_p$ increases with $z$ since the sound speed rises alongside the temperature (Fig.~\ref{nonIDEAL-Disk}).  We note that CSs even longer than $H_p$ could form above the $\beta=1$ surface, where the total pressure is mostly magnetic.  The magnetic pressure dominates above some height not just in the simplified magnetic geometry we consider here, but also in more detailed MHD treatments of protoplanetary disks' magnetically-launched winds \citep{2015ApJ...801...84G, 2020ApJ...896..126G, 2022A&A...667A..17M, 2023ASPC..534..567P}.

\subsection{{The role of AD and Hall effect in magnetic reconnection.}}
\label{ADHallRes}
The barred and dotted overlays on Fig.~\ref{Disk-Crit-Length} indicate where the AD and Hall effect, respectively, can alter the critical thresholds for fast magnetic reconnection.

{AD is relevant when there is a degree of coupling between neutrals and ions, i.e.\ in the coupled and intermediate regimes (blue and yellow in Fig.~\ref{RegTransition}).  In the decoupled regime (green), AD does not affect reconnection since neutrals are not involved in the reconnection process.  We determine the upper boundary of the barred area in Fig.~\ref{Disk-Crit-Length} using Eq.~\eqref{ADimportant}.  Reconnection on the longest available current sheet, shown by the dashed black line, is unaffected by AD at the heights plotted.  AD does however act on shorter CSs throughout the disk atmosphere.  The effects of AD on the trigger conditions for IT were treated by \citet{Zweibel:1989} and \citet{Puccietal:2020} neglecting the electron-neutral collisions in Ohm’s law (Sec.~\ref{PIPSummary}).  Extending such modeling to all three fluids -- ions, electrons, and neutrals -- would not change the maximum reconnection speed but might slightly modify $a_c$.}

For the Hall effect, the upper boundary of the dotted area in Fig.~\ref{Disk-Crit-Length} is the ion inertial length $d_i$, calculated at each height using the locally most-abundant charge carrier from Fig.~\ref{Abundance}.  The Hall effect alters the onset of fast tearing instability as discussed in \citet{Puccietal:2017}, \citet{2019ApJ...883..172S}, \citet{2021PhPl...28d2108J}, and Sec.~\ref{Sect:Sun}.  The Hall effect is destabilizing, making the CS unstable to fast tearing at slightly larger thicknesses, as quantified in \cite{Puccietal:2017}.

\subsection{Comparing disk diffusion regimes.}

{Three non-ideal terms occur in the magnetic induction equation~\ref{adimentionalind}.  Of these, only the Ohmic diffusivity is essential to the triggering of fast magnetic reconnection through the tearing instability. The AD and Hall effect change the geometry of fast-reconnecting CSs. Though Ohmic dissipation produces heating even in current sheets not undergoing tearing instability, it is much slower there in transferring energy from the magnetic field to the gas and depends on the local collision frequencies.}

{In contrast, all three non-ideal terms can be important for the magnetic field's global evolution in protoplanetary disks, affecting for example the existence and behavior of MRI turbulence \citep{2022arXiv220309821L} and the Parker magnetic buoyancy instability \citep{1996MNRAS.279..767T}.  Which non-ideal term dominates versus height in the disk atmosphere and strength of the magnetic field was explored by \citet{2011ApJ...739...50B} for a model disk with temperature independent of height, but treating some of the same ionization and recombination reactions we consider here.  There it was often true that the Ohmic diffusivity dominated near the midplane, while AD dominated for strong magnetic fields and for the atmosphere's upper layers.  Our calculation with the thermodynamics coupled to the chemistry yields a disk atmosphere much hotter than the midplane but nevertheless qualitatively similar diffusion regimes, as shown in the left panel of Fig.~\ref{Regimes-wardle-rec}.  There is no combination of field strength and height for which the Hall term dominates in this model disk annulus.}

\subsection{{Fractal and recursive reconnection.}}

\begin{figure*}
\includegraphics[width=170mm]{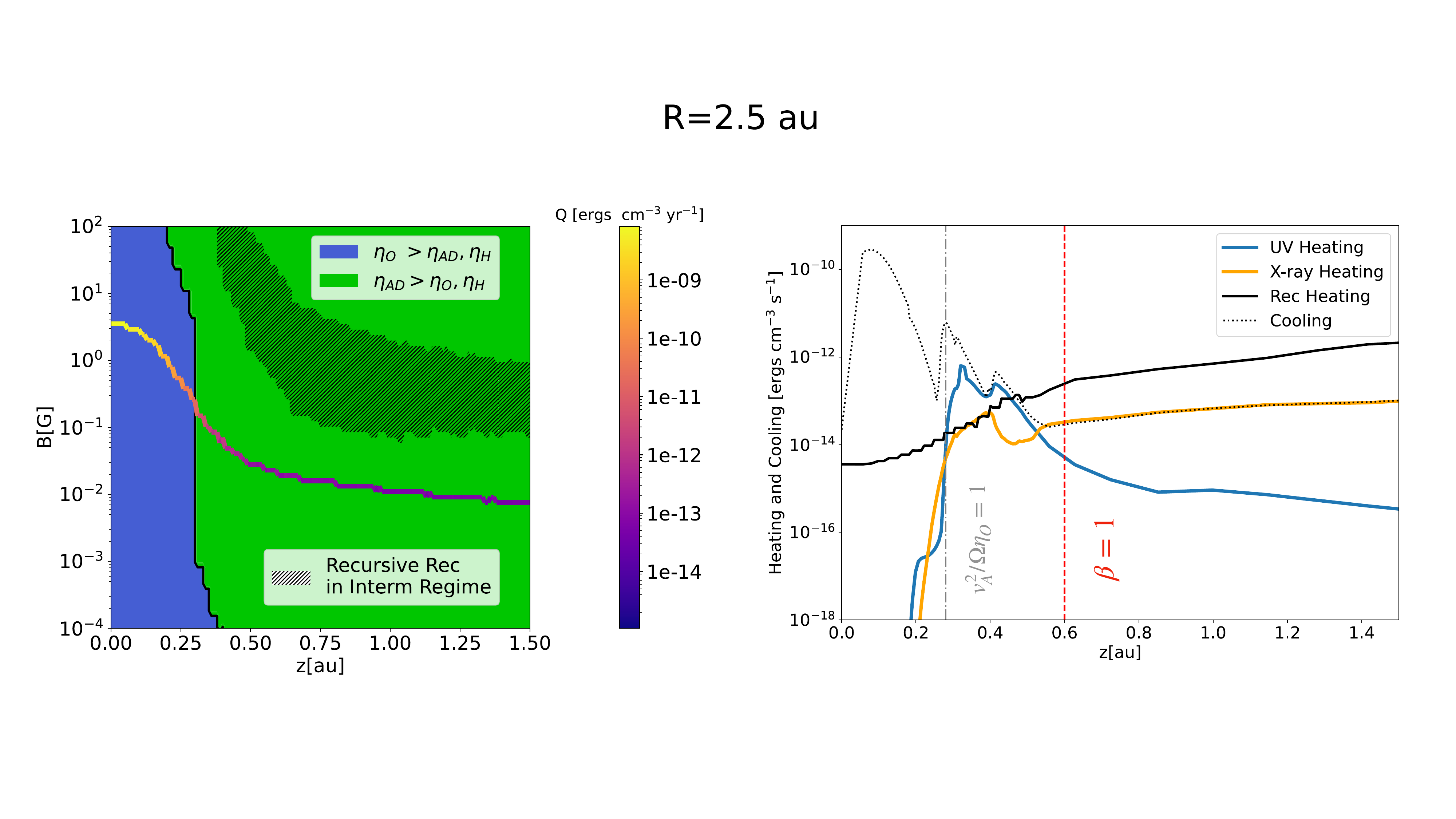}
\caption{Left panel: Largest of the magnetic diffusivities vs.\ height $z$ above the disk midplane and strength $B$ of the magnetic field.  The Ohmic term dominates in the blue and the ambipolar diffusion term in the green region.  Superimposed bars indicate where secondary reconnection events occur in the intermediate regime.  The colored curve across the center indicates the volumetric heating rate {$Q$ (erg cm$^{-3}$ yr$^{-1}$)}  where the plasma $\beta$ parameter is unity at each height $z$, using the right-hand color scale.  Right panel: Total cooling rate (dotted black curve) and selected heating rates vs.\ height $z$.  The blue curve is UV heating and the orange curve is X-ray heating.  The main heating process in the thermochemical model above 0.6~au is the X-ray heating.  Reconnection heating is stronger than the X-ray heating in the disk atmosphere, by the estimate shown as the solid black curve.  This estimate is for the coupled regime, assuming the efficiency of conversion of magnetic to heat and kinetic energy is just 10\%, and only one-third of the total field participates, so the strength of the reconnecting field is $B=10$~mG.  Since reconnection exceeds all other heating processes, it can affect the disk atmosphere's temperature, composition, and emitted radiation.}
\label{Regimes-wardle-rec}
\end{figure*}

As discussed in Sec.~\ref{cascade-fractal}, while the primary trigger occurs in the coupled regime, the secondary CS reconnection will most probably lead to regimes in which ions and neutrals are not fully coupled.
Assuming the CS length to be $H_p$ (the longest CS achievable in the system) we estimated the primary reconnection onset occurs in the fully coupled regime. In Fig.~\ref{Regimes-wardle-rec} (left) we superimposed bars when the secondary onset occurs in the intermediate regime. Shorter CSs than $L=H_p$ are not prevented to form, so that primary reconnection trigger could also occur in the intermediate or decoupled regime. Indeed, the length of the CS -- and so its thickness to achieve fast reconnection -- is affected by the local dynamics and by the CS thinning.

\subsection{Consequences for disk heating.}
\label{diskHeating}

In this section we explore whether magnetic reconnection results in plasma heating and particle acceleration capable of affecting the protoplanetary disk's temperature, chemical composition, and dynamics.

We focus on the layer where the magnetic and gas pressures are comparable.  This layer is important if MRI is present since MHD modeling of small disk patches indicates the $\beta=1$ surface separates a turbulent interior from a magnetized disk corona.  Fields generated beneath are buoyant and rise into the corona where they can undergo reconnection \citep{2000ApJ...534..398M, 2011ApJ...732L..30H}.  The $\beta\sim 1$ layer is important too if MRI is absent, since magnetocentrifugal winds are typically launched around this height \citep{1993ApJ...410..218W, 2010MNRAS.401..479K, 2014ApJ...791..137B, 2015ApJ...801...84G, 2020ApJ...896..126G, 2024arXiv240103733I}.  Reconnection heating near the launching point could change the launching speed and thus the rates at which the winds remove the disk's mass and angular momentum, with consequences for disk lifetimes.

To see whether reconnection heating is significant, we compare its rate with the other heating processes at work in the model protoplanetary disk.  The volumetric reconnection heating rate $Q\simeq B^2/(8\pi)\times(v_A/L)$ has units erg cm$^{-3}$ s$^{-1}$.  We assume the fast reconnection is triggered in the fully-coupled regime, which is a reasonable assumption for the longer CSs with $L\sim H_p$ (Fig.~\ref{Disk-Crit-Length}).

$Q$ is shown as a function of magnetic field strength in the left panel of Fig.~\ref{Regimes-wardle-rec} by color variations along the curve marking where the plasma parameter $\beta=1$.  The strength of the reconnecting field is taken to be one-third of the total field, just as for our fiducial total field strength of 30~milligauss (Sec.~\ref{ElsasserNumbers}).  The right panel of Fig.~\ref{Regimes-wardle-rec} is specific to the fiducial 30~milligauss total field strength and compares $Q$ against the main heating processes in the thermochemical disk model.  Above about $z=0.5$~au, the reconnection heating (solid black curve) exceeds all other heating processes.  Reconnection thus has the potential to raise the temperature of the gas in and around CSs.

An important question that cannot be addressed without modeling the magnetic dynamics in detail is when, for how long and where reconnection happens in such protoplanetary disk contexts.  Even MHD simulations that do not resolve the CSs may yield information on the distribution of $Q$ over space and time, useful for evaluating the contribution to heating the disk atmosphere.  However, even without such modeling we can obtain an upper limit to the time-averaged reconnection heating rate per unit disk area.  This upper limit is the rate at which the accretion flow converts the gravitational potential energy of the inward-spiraling disk gas into other forms, such as magnetic energy.

The power released by steady-state accretion per unit disk area is $D \sim (3/4\pi) {\dot M} \Omega^2$ \citep{1981ARA&A..19..137P, 2010arXiv1005.5279S}, which amounts to 400~erg cm$^{-2}$ s$^{-1}$ using a typical accretion rate onto T Tauri stars of ${\dot M} = 10^{-8}M_{\odot}$~yr$^{-1}$ and the orbital frequency $\Omega$ for our location 2.5~au from the star.  By comparison the height-integral of $Q$ from Fig.~\ref{Regimes-wardle-rec} right panel is $17$~erg cm$^{-2}$ s$^{-1}$, more than an order of magnitude below the available accretion power.
Thus reconnection can heat the disk atmosphere at rates exceeding all other known heating processes while consuming a power well within that available locally from the accretion flow.

\subsection{Reconnection Onset in Global MHD simulations.}
To quantify reconnection heating in global models of MRI turbulence and magnetocentrifugal wind launching, it will be necessary to spatially resolve the onset of fast reconnection.  We explore whether this is feasible for reconnection taking place in the layers where $\beta>1$ and starting with CS thickness $a>a_{c1}$, so that the reconnection is in the fully-coupled regime.

From the discussion in Sec.~\ref{FastRec:Summary}, the critical thickness to trigger fast reconnection $a_c \sim LS^{-1/3}(\rho_{i} / \rho_{n})^{1/6}\sim 0.008$~au, where $S\sim 10^{10}$ using the Ohmic diffusivity from $z=1$~au in the thermochemical model disk annulus.  In two-dimensional MHD simulations, grids of 1,000~cells are feasible over a domain extending from the midplane up to $z=2$~au to span the disk layers where reconnection is taking place.  The vertical grid spacing $\Delta z\sim 0.002$~au is then small enough to identify CSs capable of beginning fast reconnection.  {An MHD calculation on such a grid could estimate the places and times where reconnection produces heating at the rates $Q$ discussed above in Sec.~\ref{diskHeating}.}

\section{Conclusions}
\label{Sect:Conclusions}

We have investigated the onset of fast magnetic reconnection in the partially-ionized plasmas found in the lower atmospheric layers of the sun and in the atmospheres of protoplanetary disks.  We combined an analytic picture of the onset of reconnection via the tearing instability with models of the plasma's composition and thermodynamics in the solar and disk atmospheres.  The results indicate that reconnection may contribute to heating these environments.  The main conclusions are:
\begin{enumerate}
\item Even when reconnection begins with the ions, electrons, and neutrals all well-coupled by collisions, the tearing instability breaks the initial current sheet into secondary sheets at successively smaller spatial scales, where magnetic dissipation occurs in the intermediate or the completely decoupled regime.  As tearing proceeds recursively, the ions fully decouple from the neutrals, raising the effective Alfv\'en speed and accelerating the conversion of magnetic energy into heat.
\item In the solar photosphere and chromosphere, current sheets can have lengths up to the pressure scale height.  This provides a wide range of scales over which reconnection begins in the fully-coupled regime, so that neutrals are involved in the reconnection dynamics.  {These initial current sheets have thicknesses $a=10-100$~km, the upper end of the range being near the resolution limit of DKIST \citep{Campbelletal23}.  Since this is also about the photon mean free path at the photosphere \citep{Judge2015}, much thinner current sheets would be visible only indirectly through their heating and brightening effects.} 
\item {The Hall effect and ambipolar diffusion alter the critical scales for fast tearing instability's onset.  Corrections due to the Hall effect \citep{Puccietal:2017} and ambipolar diffusion \citep{Puccietal:2020} have been determined separately, but not together, though the Hall term likely is important only in the decoupled regime.  Ambipolar diffusion can steepen current sheets in the presence of neutral sheets with no guide field \citep{BrandenburgZweibel:1994,2020RSPSA.47690867N}, but not more generally, so that the results of \cite{Puccietal:2020} should hold, but we defer analyses of the linear stability of AD steepened sheets to future work.}
\item In protoplanetary disks, current sheets' length is limited by the density scale height when plasma $\beta>1$.  In this case there is a wide range of scales over which reconnection is triggered in the fully-coupled regime, so neutrals are also accelerated by the process. 
\item {Where the disk is well-magnetized with $\beta<1$, the largest coherent scales available can be governed by disk winds if these are not disrupted by other instabilities.  At higher plasma $\beta$ where MRI can develop, large-scale magnetic structures may also be created via channel flows.  Channel flows' breakup} by magnetic reconnection can drive time-dependent disk winds \citep{2010ApJ...718.1289S} that could move the disk surface up and down.  We have shown that in such large-scale reconnecting current sheets, the neutrals are accelerated directly by the reconnection.  The whole plasma can thus be lifted up as part of the reconnection ejecta.
\item Our work also has implications for the idea that ambipolar diffusion steepening of protoplanetary disks' midplane current sheet induces rapid accretion, pinching the field into a reversing configuration, removing magnetic flux, and yielding dense rings \citet{2018MNRAS.477.1239S}.  These processes may be greatly accelerated if the current sheet thins enough that reconnection takes place in the decoupled regime.
\item {The layer in protoplanetary disk atmospheres where the plasma $\beta$ parameter reaches unity and magnetic winds are launched is susceptible, if current sheets form, to reconnection at rates sufficient to increase the local sound speed and so the wind launching speed.} 
\item In our model disk atmosphere the power deposited locally by magnetic reconnection is sufficient to raise temperatures above those established by the stellar X-ray and UV heating, even for the minimum field strength $B\approx 10$~milligauss consistent with magnetically-driven accretion.  Stronger fields would provide even more magnetic energy for conversion to heat.  Such heating would alter the atmosphere's chemical composition and radiative emissions.  A related scenario was proposed by \citet{2013ApJ...767L...2M} with reconnection in a turbulent environment heating the parts of the protosolar disk near the young Sun to produce the glassy, spherical chondrules found in many primitive meteorites.
\end{enumerate}
Further investigation of the magnetic dissipation rates and timescales and the pathways taken by the deposited energy should include modeling that treats the multi-species nature of the plasmas.  Multi-fluid treatment of the partially ionized solar plasma has been demonstrated by \citet{SinghKrishan2010, Khomenkoetal2014, Alharbi2022}. 
Through multifluid, multispecies modeling of the solar atmosphere with hydrogen and helium, \citet{2023ApJ...946..115W} showed that the two species' differing dynamics drove chemical reactions involving helium and enriched the helium abundance in the solar wind and coronal mass ejections.  They also showed the heating rates for each species at the CS location.  Here we would like to point out that an important consequence of the fractal reconnection scenario is the eventual decoupling of the ions from the neutrals, populating energetic tails in the electrons and ions, whose effects at large Knudsen numbers \citep{Scudder92}, might be fundamental to the subsequent coronal expansion into the solar wind.  Recent observations of ubiquitous lower coronal jetting seem to provide indirect evidence for the role of such processes \citep{Raouafietal23,Bale23}.
In the context of PPDs a multifluid treatment is relevant at intermediate heights from the disk mid-plane in which heavier species than hydrogen act like the positive charge carriers, as shown by the chemical model adopted in this paper.

\acknowledgements
Copyright 2023.  All rights reserved.  We would like to thank Prof.\ Kazunari Shibata for fundamental discussions and insights on the physics of the solar atmosphere and magnetic reconnection. F.P.\ would like to thank Dr.\ Yasuhiro Hasegawa for insights on wind launching driven by magnetic reconnection.
F.P.'s research was supported by an appointment to the NASA Postdoctoral Program at the Jet Propulsion Laboratory, administered by Oak Ridge Associated Universities under contract with NASA.
K.A.P.S.\ gratefully acknowledges the UGC Faculty Recharge Program of the Ministry of Human Resource Development (MHRD), Govt.\ of India and University Grants Commission (UGC), New Delhi, the incentive grant of the Institute of Eminence (IoE) Program, BHU, and the Visiting Associateship Program of IUCAA, Pune.
M.V.\ was supported by the NASA Parker Solar Probe Observatory Scientist Grant No.\ NNX15AF34G.
M.E.I.\ acknowledges support from the German Science Foundation (Deutsche Forschungsgemeinschaft, DFG) within the Collaborative Research Center SFB1491.
This work was carried out in part at the Jet Propulsion Laboratory, California Institute of Technology, under contract with NASA and with the support of NASA's Exoplanets Research Program through grant 17-XRP17\_2-0081 to N.J.T.

\bibliography{Bib_PIP}
\bibliographystyle{aasjournal}
\end{document}